\documentclass[iop,revtex4]{emulateapj}
\newcommand{\mytilde}{\raise.19ex\hbox{$\scriptstyle\sim$}}
\usepackage{hyperref}
\usepackage{xcolor}
\begin{document}
\newcommand{\kms}{\ensuremath{\mathrm{km\,s}^{-1}}}


\title{Toward Solving the Puzzle: Dissecting the Complex Merger A521 with Multi-wavelength Data}
\author{MIJIN YOON\altaffilmark{1,2}, WONKI LEE\altaffilmark{2}, M. JAMES JEE\altaffilmark{2,3}, KYLE FINNER\altaffilmark{2}, RORY SMITH\altaffilmark{4}, AND JAE-WOO KIM\altaffilmark{4} }
\altaffiltext{1}{Ruhr-University Bochum, Astronomical Institute, German Centre for Cosmological Lensing, Universitätsstr. 150, 44801 Bochum, Germany; yoon@astro.rub.de}
\altaffiltext{2}{Department of Astronomy, Yonsei University, Yonsei-ro 50, Seoul, Korea; mjyoon@yonsei.ac.kr, mkjee@yonsei.ac.kr}
\altaffiltext{3}{Department of Physics, University of California, Davis, California, USA}
\altaffiltext{4}{Korea Astronomy and Space Science Institute, Daedeokdae-ro 776, Yuseong-gu, Daejeon 34055, Republic of Korea}

\begin{abstract}
 A521 has been a subject of extensive panchromatic studies from X-ray to radio. The cluster possesses a number of remarkable features including a bright radio relic with a steep spectrum, more than three distinct galaxy groups forming a filament, and two disturbed X-ray peaks at odds with the distant position and tilted orientation of the radio relic. These several lines of evidence indicate a complex merger. In this paper, we present a multi-wavelength study of A521 based on Subaru optical, {\it Hubble Space Telescope} infrared, {\it Chandra} X-ray, GMRT radio, and MMT optical spectroscopic observations.
Our weak-lensing (WL) analysis with improved systematics control reveals
that A521 is mainly composed of three substructures aligned in the northwest to southeast orientation. These WL mass substructures are remarkably well-aligned with the cluster optical luminosity distribution constructed from our new enhanced cluster member catalog. These individual substructure masses are determined by simultaneously fitting three NFW profiles. We find that the total mass of A521 modeled by the superposition of the three halos is $13.0_{-1.3}^{+1.0} \times 10^{14}M_{\sun}$, a factor of two higher than the previous WL measurement.
With these WL mass constraints combined with X-ray and radio features, we consider two merging scenarios, carry out the corresponding numerical simulations, and discuss strengths and weaknesses of each case. 
\end{abstract}

\section{Introduction}

Cluster collisions happen at the end of the merger chains in the hierarchical structure formation paradigm  \citep[e.g.,][]{ 2005Natur.435..629S,2014MNRAS.444.1518V}. In-depth studies of these cosmic events provide unique opportunities to understand the growth of the large scale structure of the universe. In addition, more recently, these merging clusters have been regarded as 
useful laboratories, enabling particle experiments beyond the energy scale executable on Earth. One of the most outstanding questions that the experiments can address is the nature of dark matter.
However, we do not have any control over the setup of the experiments (i.e., initial conditions prior to collision). Moreover, only a single snapshot
of the long merger history is available to us.
Therefore, careful characterization of the current status of the merging clusters and robust reconstruction of merger scenarios
must be carried out in order to
fully utilize these tremendous opportunities to address the fundamental questions in physics.

Radio relic clusters are a rare class of merging clusters that exhibit large-scale ($\gtrsim 1~$Mpc) synchrotron emissions in the cluster outskirts \citep[e.g.,][]{2001A&A...376..803G,2009A&A...503..707B,2010Sci...330..347V,2011ApJ...736L...8B,2018MNRAS.477..957D,2020A&A...633A..59P}. 
These large-scale radio features, so-called radio relics, are believed to trace merger shocks in the intracluster medium (ICM), which are created at the core impact and propagate outward along the merger axis. Thus, a radio relic observation informs us of the merger details including the
time since the impact and the orientation of the merger axis, which are invaluable pieces of information in our reconstruction of the merger scenario.

We, Merging Cluster Collaboration (MC$^2$)\footnote{\url{http://www.mergingclustercollaboration.org}}, are conducting a series of studies to investigate radio relic clusters \citep[e.g.][]{ 2019ApJ...882...69G,2019ApJS..240...39G,2020ApJ...894...60L}. Our short-term scientific goals are to robustly measure the distributions of the three cluster constituents (i.e., gas, dark matter, and galaxies) for our sample of radio relic clusters based on multi-wavelength observations and constrain most probable merger scenarios through iterative numerical simulations. The long-term goals are to address the particle acceleration inefficiency problems and measure the self-interaction cross section of dark matter as a step to reveal its fundamental properties. 

In this paper, we present a multi-wavelength study of the complex cluster merger
A521 at $z=0.247$. The cluster has been a subject of extensive multi-wavelength studies.
Based on their analysis of optical and X-ray observations, \cite{2000A&A...355..461A} and \cite{2000A&A...355..848M} reported that A521 is rich in substructure. 
The ROSAT/HRI image analysis of \cite{2000A&A...355..461A} revealed two X-ray peaks separated by $\mytilde0.5$~Mpc in the NW-SE orientation.
However, this rather simple binary structure in X-ray 
is in contrast with a much more complex structure in optical \citep{2000A&A...355..848M, 2000A&A...355..461A}.
\cite{2000A&A...355..461A} hypothesized that A521 might be located at the crossing of two filaments and this complex structure in the galaxy distribution
might arise from two independent mergers at different epochs.

The two early studies of \cite{2000A&A...355..848M} and \cite{2000A&A...355..461A} were based on
their 41 spectroscopic members.
\cite{2003A&A...399..813F} nearly tripled the number of the spectroscopic members (from 41 to 113) and confirmed that the A521 galaxy distribution indeed possesses a complex structure. The authors identified seven groups forming the NW-SE filament and one perpendicular (NE-SW) ridge in the cluster galaxy distribution. In addition, their velocity analysis indicates that A521 is also complex in the line-of-sight (LOS) direction. If these substructures are real as claimed by the authors, we may be witnessing multiple merging events in A521.

Studies with improved X-ray observations \citep{2006A&A...446..417F,Bourdin_2013} using \textit{Chandra} and XMM-\textit{Newton} indicate that perhaps the X-ray data too may support the complexity of the cluster substructures and mergers proposed by optical studies. In particular,
\cite{Bourdin_2013} showed that A521 possesses interesting surface brightness and temperature features including two cold fronts and two shock fronts. 

The A521 radio relic image was first presented by \cite{2000A&A...355..461A} with the archival NVSS data. Even with the broad beam ($\mytilde45\arcsec$), the arc-like radio feature is clear in the NVSS image, although this radio morphology was neither explicitly mentioned nor referred to as a ``relic" in \cite{2000A&A...355..461A}.
\cite{2003A&A...399..813F} and \cite{2008A&A...486..347G} confirmed the presence of the relic with
higher-resolution Very Large Array (VLA) 1.4 GHz and Giant Metrewave Radio Telescope (GMRT) 610MHz observations, respectively.

Despite the above wealth of information available for A521, no convincing merging scenario that explains the current multi-wavelength 
observations has been presented. Given the large number of substructures seen in the cluster galaxy \citep{2003A&A...399..813F} 
and ICM \citep{Bourdin_2013} distributions,
it is difficult to suggest plausible merging scenarios that account for the entire
complexity.

We argue that one powerful method that helps to reduce the dimension of the complexity in the reconstruction of the A521 merger scenario is weak-lensing (WL) analysis. WL studies enable us to identify the most significant cluster substructures in mass. This in turn allows us to
identify apparent galaxy over-densities, that are not real, and simply due to projection effect. Because the mass-to-light ratio (M/L) values in bona fide galaxy groups and clusters are at least an order of magnitude higher than in individual galaxies, in general a chance projection of multiple isolated galaxies along the LOS direction does not lead to a significant WL detection. However, needless to say, great caution must be used in WL studies, which sometimes produce noise peaks due to chance alignments or poor systematics control.

Therefore, the primary goal of our study is to constrain the detailed mass distribution in A521 with a careful WL analysis. Although A521 is one of the 30 LoCuSS clusters \citep{2019MNRAS.484...60M} studied with WL in \cite{Okabe_2010}, the authors 
do not discuss its substructures.
In our study, we characterize the substructure properties (e.g., mass, position, centroid uncertainty, etc.) with WL and provide detailed comparison with our multi-wavelength data. In particular, we enhance our knowledge on the galaxy distribution with our new spectroscopic observations using MMT/Hectospec, which adds 67 new cluster members. 
Finally, we discuss the A521 merging scenario with our hydrodynamic simulations.

We present our paper as follows. In \textsection\ref{sec:observation}, we introduce the data from Subaru/Suprime-Cam, {\it Hubble Space Telescope}/Wide Field Camera 3 ({\it HST}/WFC3), MMT/Hectospec, {\it Chandra}, and GMRT observations. In \textsection\ref{sec:analysis}, we explain detailed steps in our WL analysis, which include member and source selections, point-spread-function (PSF) modeling, shape measurement, and the cluster mass estimation. Discussions including our merging scenarios are presented in \textsection\ref{sec:discussions} before we conclude in \textsection\ref{sec:conclusions}.

Throughout the paper, $M_{500}$ ($M_{200}$) corresponds
to the total mass enclosed within a radius, inside which the
mean density is equal to 500 (200) times the critical density of the universe
at the cluster redshift $z=0.247$ in the adopted cosmology.
We assume a flat $\Lambda$CDM universe with $\Omega_m=0.3$ and 
$H_0=70~\mbox{km}~\mbox{s}^{-1} \mbox{Mpc}^{-1}$, for which
the plate scale is $\mytilde232$~kpc~arcmin$^{-1}$ at the cluster redshift.
All quoted uncertainties are at the 1~$\sigma$ ($\mytilde68.3$\%) level.

\section{Observation}
\label{sec:observation}
\begin{figure*}
    \centering
    \includegraphics[trim=2cm 0 0 -1cm, width = 1.0\textwidth]{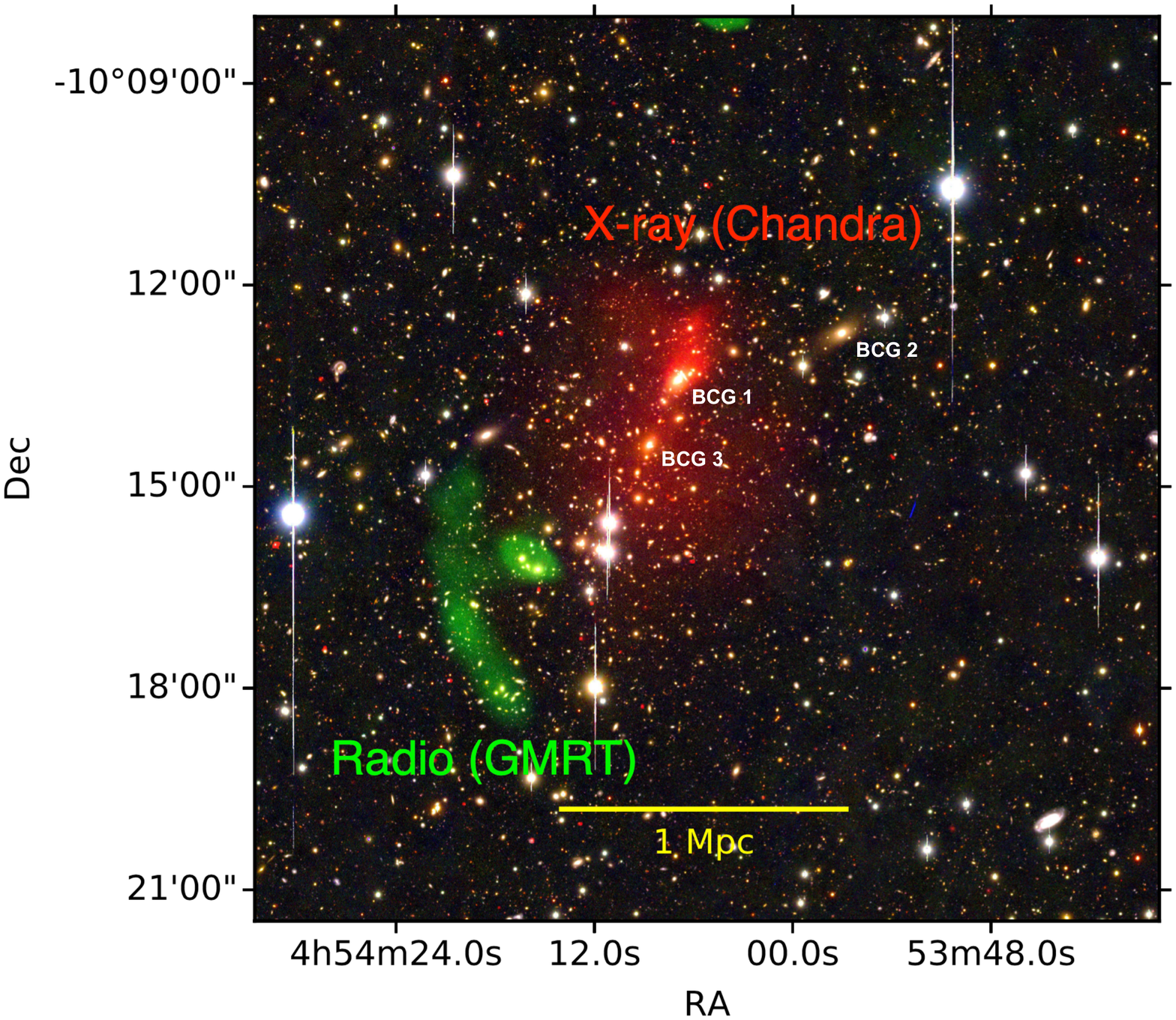}
    \caption{A521 observations with Subaru, GMRT, and {\it Chandra}.
    We use the Subaru/SuprimeCam $i'$, $R$, and $V$ images to create the color composite image. Diffuse red and green emissions show the adaptively smoothed {\it Chandra} X-ray and the 153~MHz GMRT radio images, respectively. We show the central $13.3^2$ arcmin$^2$ region (corresponding to the $3.1^2$ Mpc$^2$ physical area at the cluster redshift for the adopted cosmology). 
    The $\mytilde1$~Mpc radio relic is located in the southeastern periphery. We mark the locations of the first, second, and third BCGs for reference in the text. 
    Based on the X-ray emission, A521 is characterized by the two gas clumps aligned in the north-south orientation with the NW and SE clumps hosting the first and third BCGs, respectively. 
    The $\mytilde1$~Mpc radio relic is located in the southeastern periphery. 
     }
    
    \label{fig:color_xray_radio_map}
\end{figure*}
\subsection{Subaru/Suprime-Cam}

Subaru/SuprimeCam observations on A521 were conducted with $V$, $R$, and $i'$ on 2001 October 15, with NB816 (narrow band) on 2002 February 15, with $z'$ on 2002 February 16, and with $B$ and $U$ on 2003 February 4. 
We retrieved the datasets for the $V$, $R$, and $i'$ filters from \texttt{SMOKA}\footnote{\url{https://smoka.nao.ac.jp}}. 
The $V$, $R$, and $i'$ exposure times are 1,800~s, 1,620~s, and 2,040~s with the seeings $0\farcs59$, $0\farcs65$, and $0\farcs59$ , respectively.  We note that the previous WL analysis of \cite{Okabe_2010}  used the $i'$ and $V$ images whose exposure times are 1,320~s each. 

We used \texttt{SDFRED1} \citep{2002AJ....123...66Y, 2004ApJ...611..660O} 
to perform
the initial CCD-level data reduction (i.e. overscan \& bias subtraction, bad-pixel masking, flat fielding,  distortion correction, and AG probe masking).
The next data reduction steps including astrometric calibration and image stacking
are crucial for controlling WL systematics. While \cite{Okabe_2010} relied on \texttt{SDFRED1} for these tasks, we used {\tt SCAMP}\footnote{\url{https://www.astromatic.net/software/scamp}} \citep{2006ASPC..351..112B} and {\tt SWARP}\footnote{\url{https://www.astromatic.net/software/swarp}} \citep{2002ASPC..281..228B}.
Image processing through the combination of both software tools
has been extensively tested in the WL community \citep[e.g.,][]{2013ApJ...765...74J,2017ApJ...851...46F, 2018A&A...610A..85S, 2019A&A...625A...2K}. {\tt SWARP}
provides individual {\tt RESAMP} images, which we utilize to model the PSF variation for each CCD and epoch. Since the Subaru/Suprime-Cam PSF is epoch- and CCD-dependent, as is true with all other telescopes, this per-epoch CCD-level PSF modeling is important to address the PSF-induced galaxy shape distortion.

Objects were detected by running {\tt SExtractor} \citep{1996A&AS..117..393B} in dual-image mode and finding
at least five connected pixels whose rms values are 1.5 times the background rms.
Since the $i'$ image is deepest, we used it as the detection image. Using the same detection image for every filter allows us to maintain consistent isophotal areas and object IDs across different filters, which is important for robust color measurement. Throughout the paper, we use {\tt SExtractor}'s {\tt MAG\_ISO}  to compute object colors while {\tt MAG\_AUTO} is employed for the rest.

\subsection{{\it HST}/WFC3}
A small ($\mytilde2 \farcm 3 \times2\farcm 1$) region of the A521 field was observed with {\it HST}/WFC3 during the 2018 April 10-13 period
in F390W, F105W, and F160W with integrations of 2468 s, 2612 s, and 5223 s, respectively (PROP ID: 15435). 
The observation was done with a single pointing covering the first and third BCGs (see Figure~\ref{fig:color_xray_radio_map} for the BCG locations).
We retrieved the IR channel data (F105W and F160W) from the MAST\footnote{https://mast.stsci.edu} and processed the {\tt FLT} images following the procedures described in \cite{2017ApJ...847..117J} and \cite{2020ApJ...893...10F}.
In brief, we estimate shifts between exposures using common astronomical sources while applying the time-dependent geometric distortion correction.
The final mosaic images were created with an output pixel scale of $0\farcs05$, the Gaussian ``drizzle" kernel, and the {\tt pixfrac=0.7} parameter setting.
We measure galaxy shapes from the F160W image since its exposure
time is twice as long as that of F105W, which we use only to create a color-composite image for the blue channel.

\subsection{MMT/Hectospec}
Fiber spectrograph imaging of A521 was completed as part of a merging cluster program (PI: Finner) with the MMT Hectospec. The observations were conducted on the nights of 22 and 23 October 2019 under clear skies with seeing of $0\farcs9$ and $0\farcs7$, respectively. Hectospec consists of 300 fibers within its 1$^\circ$ diameter field of view. We utilized the 270 grating (spectral range 3650 - 9200 \textup{\AA}) and performed two configurations of three 1200~s integrations. 

Target selection was done with the Subaru data within $\mytilde40\arcmin$ of the first BCG and the Pan-STARRS \citep{2012ApJ...750...99T} DR2 data for the outer region using a $V - i'$ color versus $i'$ magnitude diagram. A linear fit was performed to the existing spectroscopically confirmed cluster members and a region $\pm0.12$ in color and $i'_{\rm BCG} < i' < 20.5$ in magnitude were passed on to the \texttt{xfitfibs} software to create the two configurations while using caution to avoid fiber collisions. Raw spectra were processed with the \texttt{hs\textunderscore pipeline\textunderscore wrap} command in HSRED 2.0 to produce sky-subtracted and variance-weighted spectra. The output spectra were then passed through the IDL task \texttt{hs\textunderscore reduce1d} where spectral template comparison was performed and redshifts were determined. To finalize the catalog, we applied the goodness of fit flag from the MMT pipeline, `zwarning' = 0. Also, we applied a cut to exclude the objects outside of the redshift window, $0.01< z < 1$. 
Our spectroscopic redshift catalog from our MMT/Hectospec observation is shown in
Table~ \ref{Tab:MMT_cat}.  

\begin{table}[t]

\centering
\caption{MMT Hectospec spectroscopic redshift catalog of the A521 field}
\begin{tabular}{llll}

\hline\hline
RA & Dec & z & $\sigma_z$ [$\times 10^{-5}$]\\
\hline
73.6481667 & -10.122686 & 0.246442 & 4.66  \\
73.3250208 & -10.179791 & 0.246458 & 1.52  \\
73.3249708 & -10.160116 & 0.246566 & 2.11  \\
73.1223458 & -10.370778 & 0.246596 & 5.32  \\
73.6332167 & -10.173622 & 0.246626 & 11.0  \\
73.3422167 & -10.346572 & 0.246646 & 3.55  \\
73.6228542 & -10.244136 & 0.246673 & 2.97  \\
73.5940167 & -10.159275 & 0.246675 & 4.94 \\
73.7117292 & -10.303531 & 0.246736 & 4.06  \\
73.7615792 & -10.313891 & 0.246775 & 5.03  \\
73.5359125 & -10.131342 & 0.246788 & 3.81 \\
73.4624292 & -10.268331 & 0.246830 & 4.81 \\
73.5984917 & -9.8299531 & 0.246849 & 2.78 \\
73.4350208 & -10.195231 & 0.246849 & 4.22 \\
\hline\hline
\end{tabular}
\tablecomments{The catalog is available in its entirety in the electronic version of this paper.}
\label{Tab:MMT_cat}
\end{table}


\subsection{Chandra}
A521 is one of the clusters that \textit{Chandra} observed during the first cycle (PI: M. Arnaud) with ACIS-I and ACIS-S. The exposure time for each instrument is $\mytilde$40~ks. The cluster was re-observed with a total exposure of $\mytilde$90 ks in Cycle 12 using ACIS-I (PI: M. Markevitch).
We retrieved both datasets from the {\it Chandra} archive\footnote{\url{https://cda.harvard.edu}} and created a deep (a total exposure of $\mytilde130$ ks) stack for ACIS-I. 
We used the {\tt csmooth} package to adaptively smooth (with the minimum and maximum significances of 3 and 5, respectively) the exposure-uncorrected image output by the {\tt merge\_obs} script. 
The resulting smoothing scale map was utilized to apply the same adaptive smoothing scheme once again to the exposure map. The final adaptive smoothing map was obtained by dividing the first {\tt csmooth} output by the second {\tt csmooth} output.
In this paper, we use the {\it Chandra} data only to compare the spatial distribution of the X-ray emission with the cluster galaxy and mass distributions and perform no independent spectral analysis. Where necessary, we quote spectral measurements found from the literature.

\subsection{Giant Metrewave Radio Telescope}

GMRT observed A521 at 153 MHz in August 2009 with integration of 10 hours. The reduction and analysis of the data were presented in \cite{Macario_2013}, who kindly provided the final reduced image to us. Readers are referred to their paper for details. The authors reported that after rigorous cleaning the total data loss was $\mytilde45\%$, which reduces the effective exposure time to less than 5 hours. Nevertheless, the resulting final radio image reveals a clear giant relic at the eastern periphery of A521.

Figure~\ref{fig:color_xray_radio_map} displays the color-composite image created with the Subaru $V$, $R$, and i' filters representing intensities in blue, green, and red, respectively. Diffuse red and green emissions show the adaptively smoothed {\it Chandra} X-ray and the 153~MHz GMRT radio images, respectively.

\section{Analysis}
\label{sec:analysis}
\subsection{Spectroscopic Cluster Member Selection}
\label{sec:spec_selection}
We merge the two publicly available spectroscopic catalogs of \cite{2000A&A...355..848M} and \cite{2003A&A...399..813F} with ours obtained from the MMT Hectospec observation of the A521 field.
We retrieved 47 objects in the \cite{2000A&A...355..848M}
study from the SIMBAD astronomical database\footnote{http://simbad.u-strasbg.fr/simbad} 
(we found that the column labels of the SIMBAD catalog were in error at the time of this writing and used the column incorrectly labeled as ``radial velocity" for $cz$). \cite{2003A&A...399..813F} added 191 new objects, which we 
downloaded from the VizieR Astronomical Server\footnote{vizier.u-strasbg.fr}.
With MMT/Hectospec, we obtained a total of 398 spectra whose redshift errors are less than $10^{-4}$ in the A521 field. Out of these 398 spectra, 15 objects are in the aforementioned public catalogs and used for cross-checking. The comparison shows that only one object has a large discrepancy ($\mytilde0.1$) and our MMT/Hectospec measurement reveals that this object is in fact a cluster member. For these 15 objects, we used our MMT/Hectospec values.

\begin{figure}
    \centering
    \includegraphics[width = 0.49\textwidth]{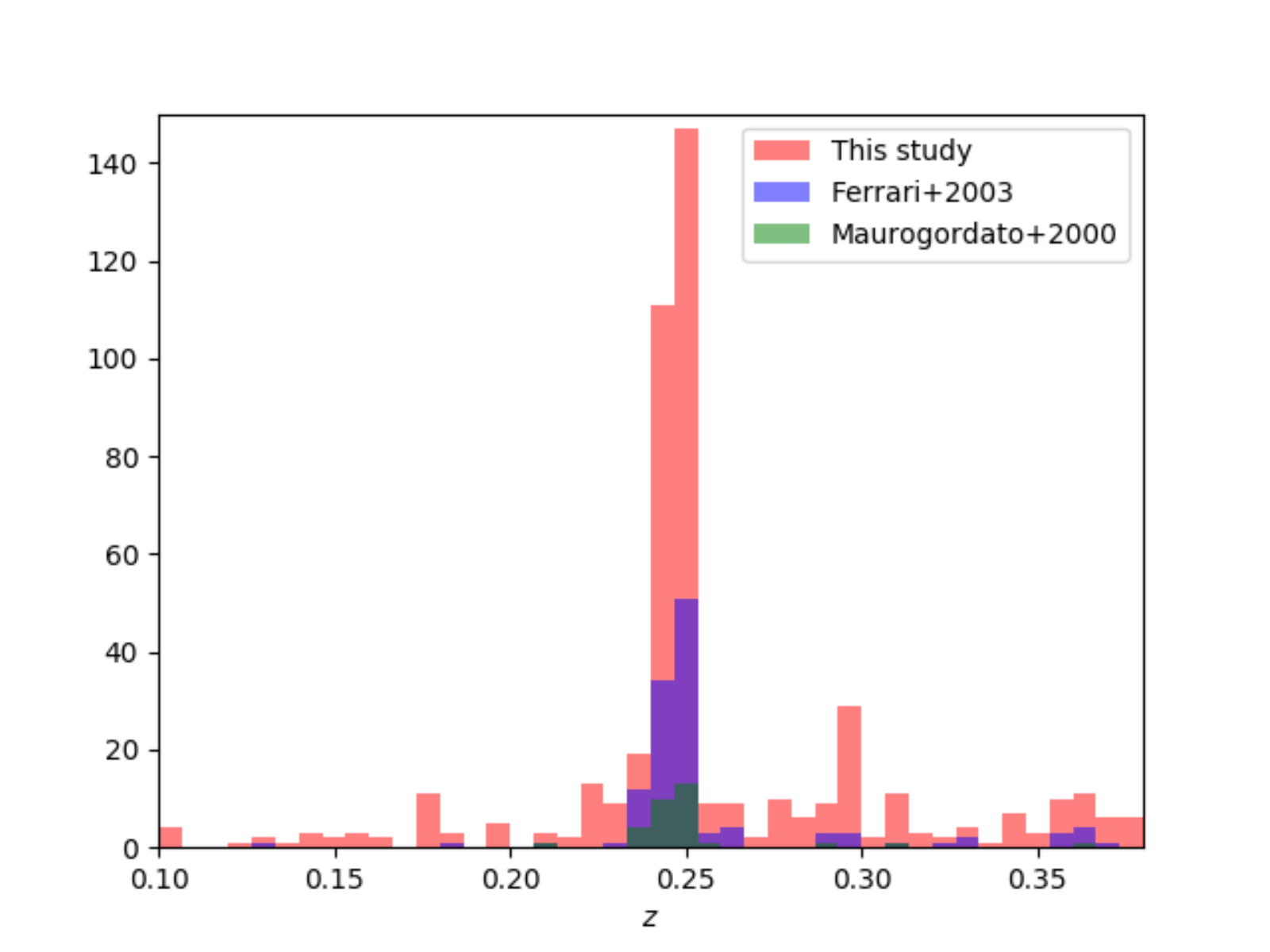}
    \includegraphics[width = 0.49\textwidth]{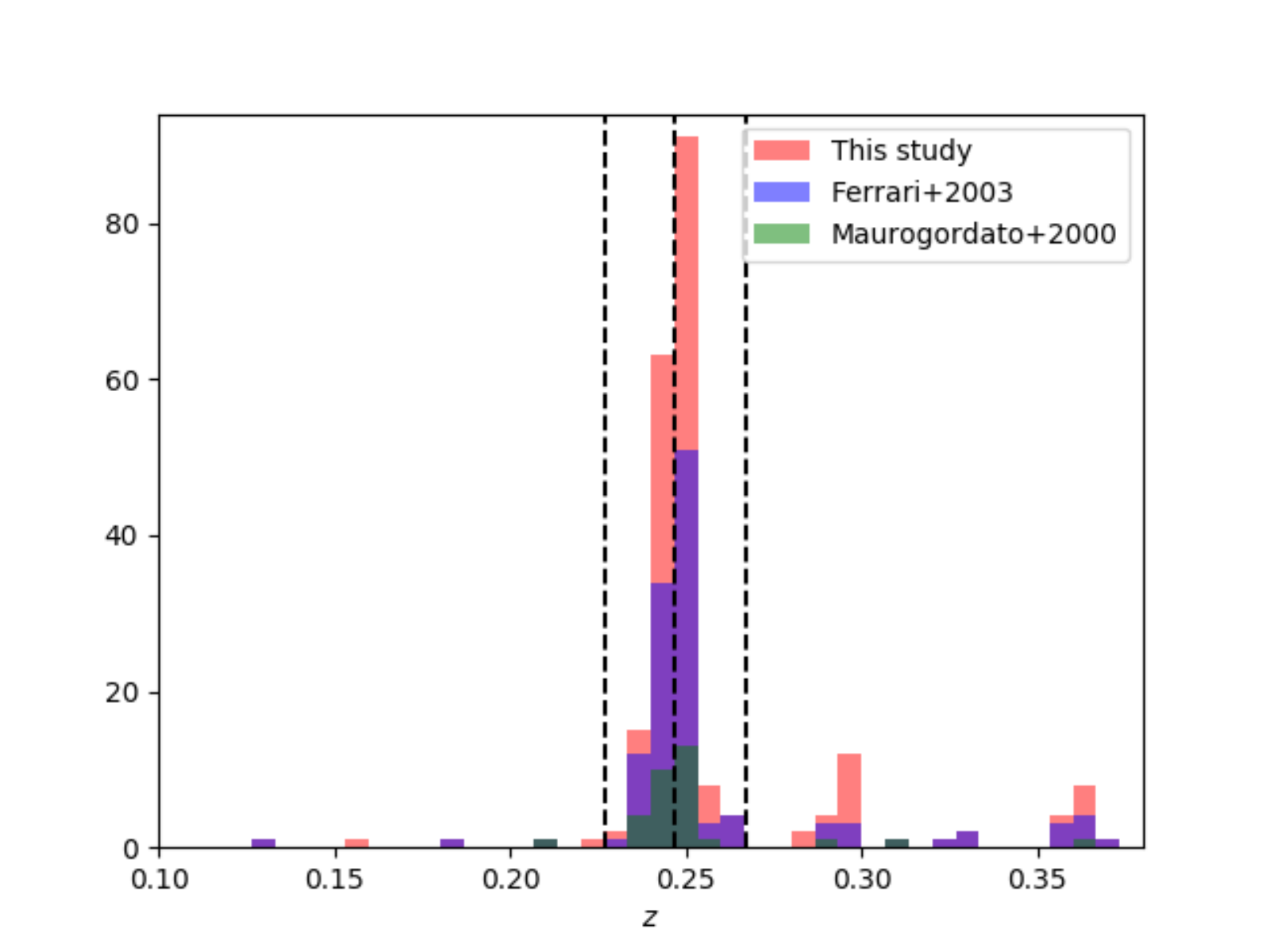}
    \caption{Redshift distribution of galaxies in the A521 field.
    We merge the two publicly available spectroscopic catalogs of \cite{2000A&A...355..848M} and \cite{2003A&A...399..813F} with ours obtained from the MMT/Hectospec observation. 
    The objects observed again are counted as new data.
    Top: We display the distribution of all sources in the A521 field covered by the large ($\mytilde1\degr$) field of view of the Hectospec instrument.
    Bottom: Same as the top except that the histogram is constructed using the objects located within the 2~Mpc radius from the first BCG. 
    The central vertical dashed line represents the
    cluster redshift determined by the biweight estimator (see text). The other two 
    vertical dashed lines mark the $\Delta z=0.02 $ interval, with which we define the A521 membership. The total number of the spectroscopic  members is 183, including the previously confirmed 116 members.}
    \label{fig:redshift_distribution_spec_z_sample}
\end{figure}

MMT/Hectospec covers a large ($\mytilde1\degr\times \mytilde1\degr$) area. We limit our analysis to the objects located within
the $r=2$ Mpc radius (this roughly corresponds to the virial radius of A521, see \textsection\ref{sec:multi_halo_fitting}) from the first BCG.
Using the biweight estimators \citep{1990AJ....100...32B}, we determined the central redshift and velocity dispersion of A521 to be $z=0.24672\pm0.00037$ and $\sigma_{1D} = 1587\pm176~\mbox{km}\mbox{s}^{-1}$, respectively, where the uncertainties were measured from bootstrap resampling.  
We selected 183 objects within $\Delta z=0.02$ (corresponding to $\mytilde3\sigma_{1D}$) from the central redshift as the A521 member galaxies.
Figure~\ref{fig:redshift_distribution_spec_z_sample} displays the redshift distribution of galaxies in the A521 field from the previous and our observations.

\subsection{Photometric Cluster Member Selection}
\label{sec:phot_cluster_member}
\begin{figure}
    \centering
    \includegraphics[width = 0.49\textwidth]{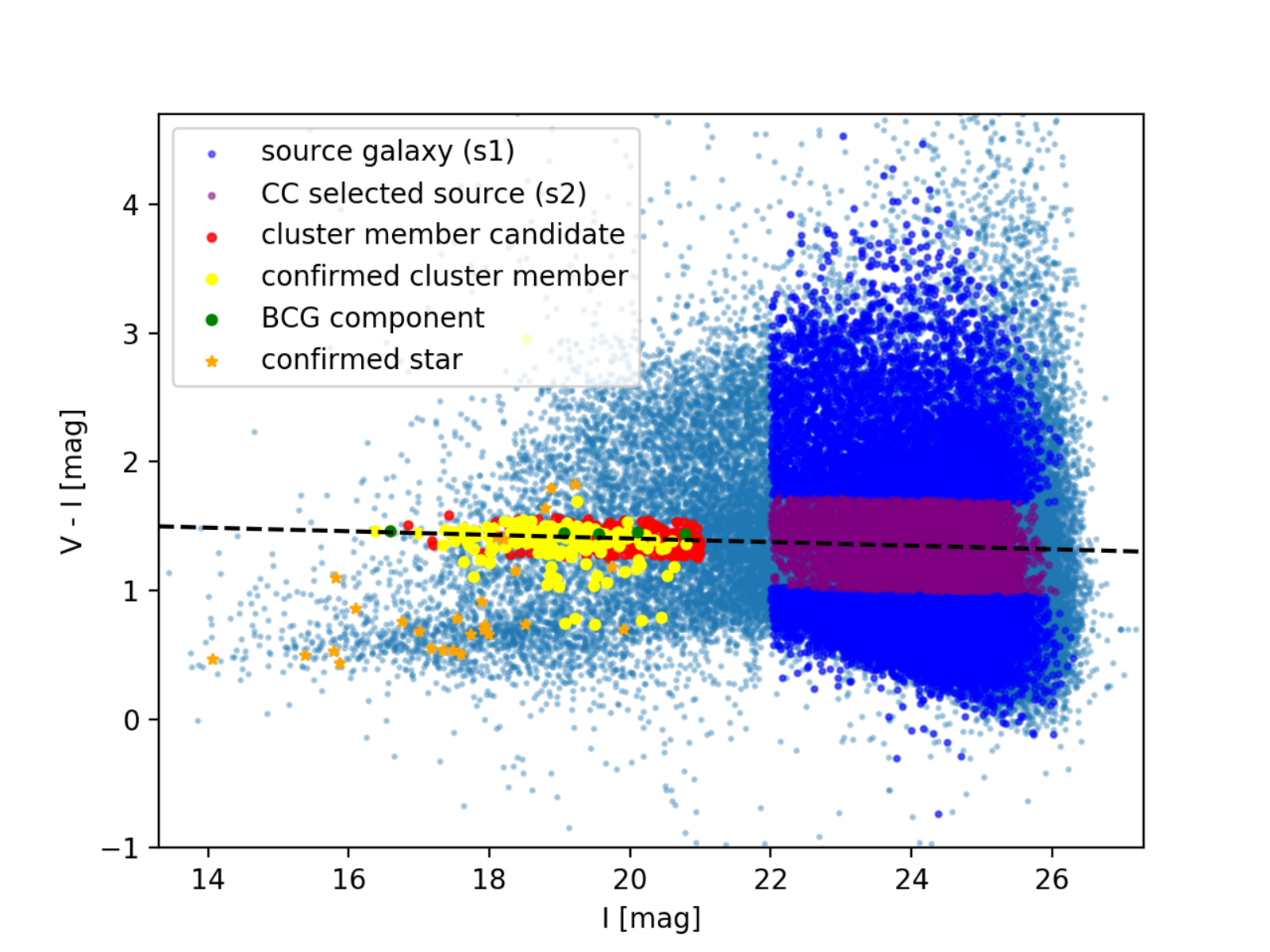}
    \includegraphics[width = 0.49\textwidth]{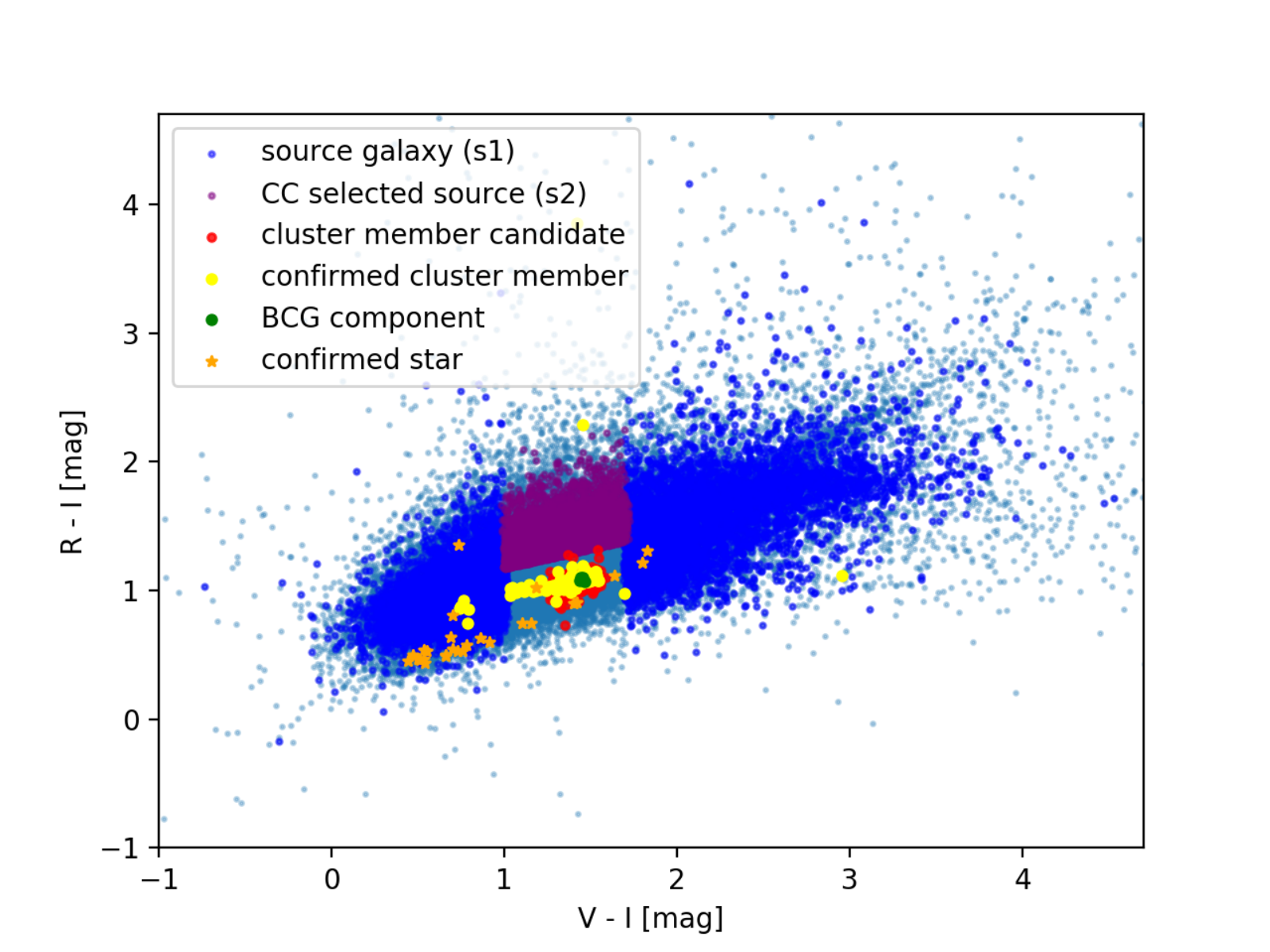}
    \caption{Color-magnitude and color-color relations in the A521 field. 
    Top: We display the $V - i'$ color versus $i'$ magnitude relation. The spectroscopically confirmed members occupy a narrow locus.
    The dashed line is the best-fit relation observed by these spectroscopic members with exclusion of some blue members: $V - i' = -0.0125i' + 1.656$. The BCG components (green) are the galaxies belonging to the BCG group hosting the first BCG \citep{2000A&A...355..848M}.
    Bottom: The $V-i'$ versus $R-i'$ colors are shown. The spectroscopic members span a narrow range in $V-i'$ color as well. 
    The best-fit relation between the two colors is given by the equation:
    $R - i' = 0.298 (V - i')+ 0.671$.
    We select the photometric cluster members and lens sources utilizing both relations.
    }
    \label{fig:cmd_for_selection}
\end{figure}

\begin{figure*}
    \centering
    \includegraphics[trim=2cm 0cm 1.2cm 1.5cm, width = 0.495\textwidth]{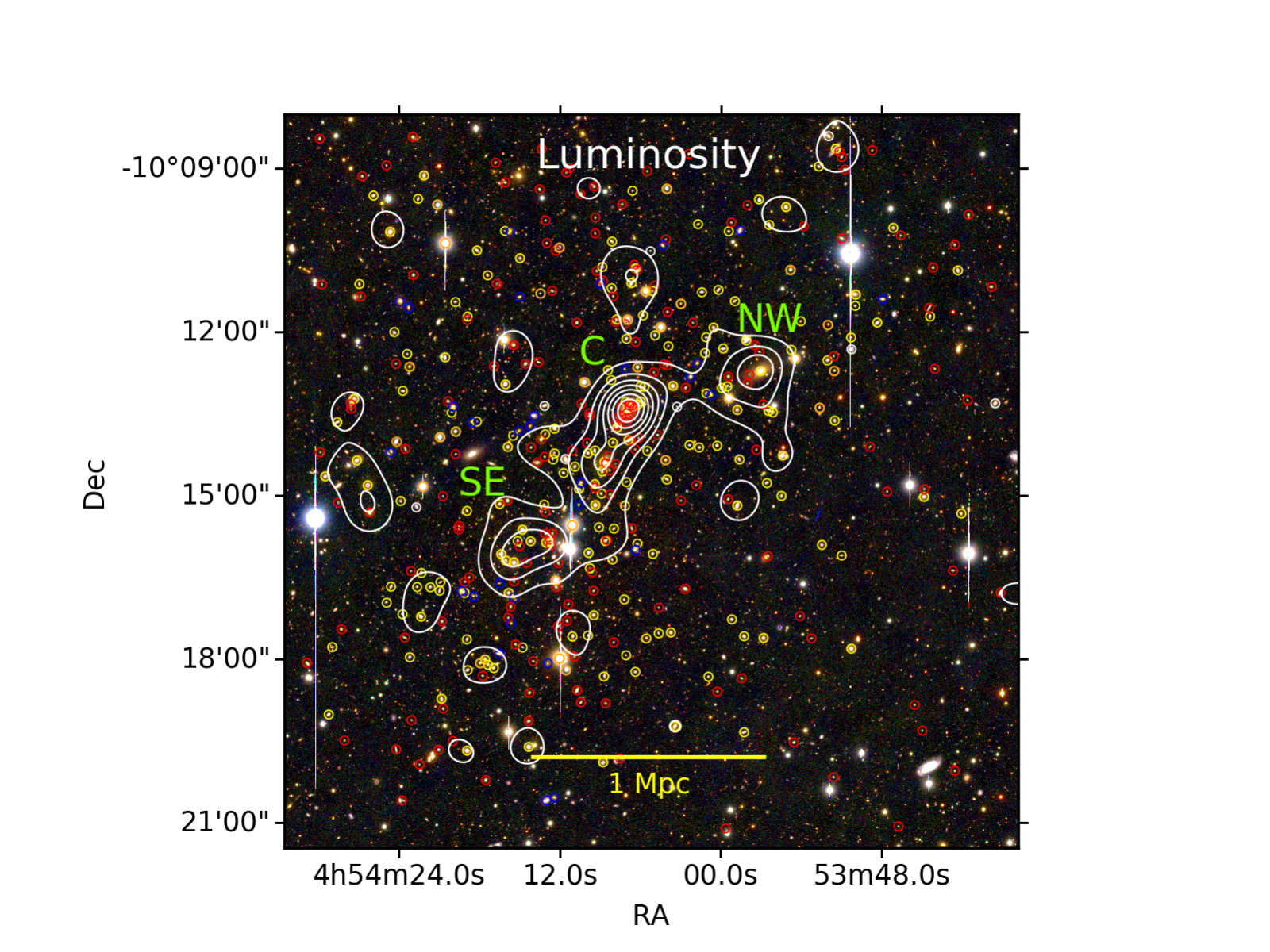}
    \includegraphics[trim=2cm 0cm 1.2cm 1.5cm, width = 0.495\textwidth]{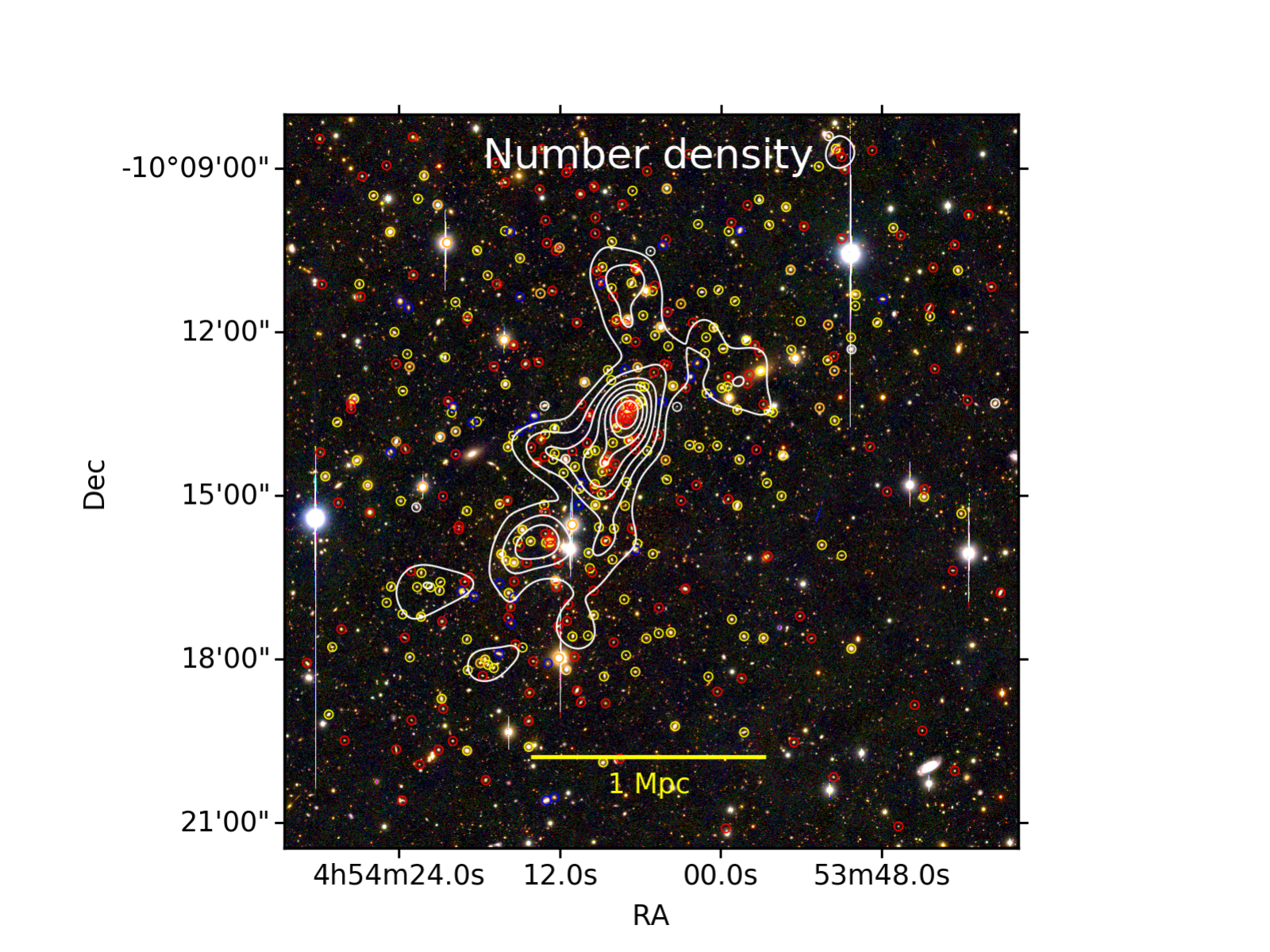}
    \caption{Cluster members identified by spectroscopic observation and photometric selection based on the color-magnitude diagram. The contours represent the luminosity-weighted  (left) and number densities (right). The luminosity-weighted (number density) contours are linearly spaced with the lowest level corresponding to 15\% (22\%) of the maximum. The circles show the confirmed members (yellow), cluster member candidates (red), confirmed foreground galaxies (white), confirmed background galaxies (blue), and confirmed stars (orange). The cyan rectangle in the right panel marks the approximate location and orientation of the ``ridge" feature claimed by \cite{2003A&A...399..813F}.
    The galaxy distributions show that A521 is mainly comprised of the three substructures: NW, C, and SE, although the NW substructure is somewhat less clear in the number density.}
    \label{fig:member_galaxy_with_img}
\end{figure*}

The spectroscopic members discussed in \textsection\ref{sec:spec_selection} occupy a tight locus
in the color-magnitude diagram, as shown in Figure~\ref{fig:cmd_for_selection}, which we utilize to define our photometric cluster members.
Our criteria are as follows. The photometric members should be sufficiently bright ($i' < $21) with reasonable magnitude measurement errors ($\sigma_i' < 0.3$) and sizes ($r_h > 1.6$).
Their $V - i'$ colors are required to be within 0.15 mag from the best-fit color-magnitude relation while their $R - i'$ colors should be within 0.2 mag from the best-fit color-color relation. We summarize the member selection criteria in Table~\ref{tab:member_gal_selection}.
Using the criteria, we select 244 galaxies as member candidates, in addition to the spectroscopic members, within the 2~Mpc radius from the first BCG. In Figure \ref{fig:member_galaxy_with_img}, we display the locations of the cluster members and member candidates on the Subaru color-composite image. The 
luminosity-weighted density map shows that A521 consists of three distinctive clumps (we denote them as NW, C, and SE in the left panel of Figure~\ref{fig:member_galaxy_with_img}), which are approximately colinear along the NW-SE direction. 
These substructures also appear in the number density map while the NW clump is somewhat less concentrated. 

This simple linear structure is in contrast with the previous claim of \cite{2003A&A...399..813F}, who identified seven groups and one ``ridge" (see the cyan rectangle in the right panel of Figure~\ref{fig:member_galaxy_with_img} for the approximate location of the ridge), which together form a cross-like structure.
Although the ``ridge" structure is hinted at by our number density map, the feature is much weaker than the NW-SE linear structure (in \textsection\ref{sec:mass_reconstruction}, we show that our WL mass structure of A521 is highly consistent with the current galaxy distribution).

\begin{table}[h]\centering
\caption{Photometric cluster member selection criteria}
\scriptsize
\begin{tabular}{cc}
\hline
\hline
Magnitude &  $i' < 21 $\\
Magnitude error & $\sigma_{i'} < 0.3$ \\ 
Color-magnitude cut & $\left| k_1 - 1.656 \right | < 0.15 $ \\ 
Color-color cut & $ k_2  < 0.2$ \\
Galaxy size &  $r_h >$ 1.6 pixel\\
\hline
\hline
\end{tabular}
\label{tab:member_gal_selection}
\tablecomments{We define $k_1$ and $k_2$ as $k_1  \equiv V - i'+0.0125i'$ and 
$k_2  \equiv R - i' -0.298(V - i')-0.671$, respectively. }
\end{table}

\subsection{PSF Modeling}

\begin{figure*}
    \centering
    \includegraphics[trim=0.5cm 0.5cm -1.5cm 0cm, width = 0.48\textwidth]{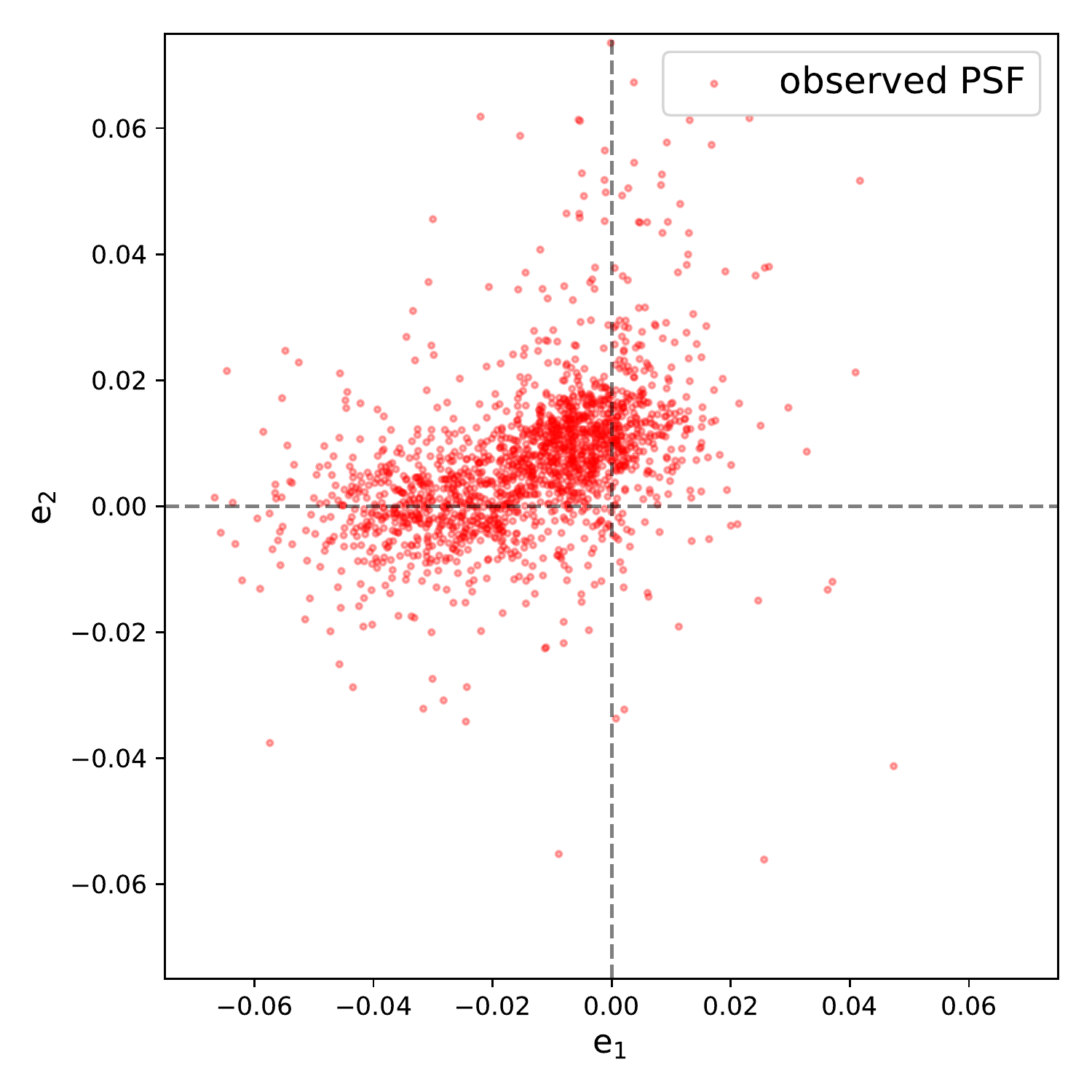}
    \includegraphics[trim=0.5cm 0.5cm -1.5cm 0cm, width = 0.48\textwidth]{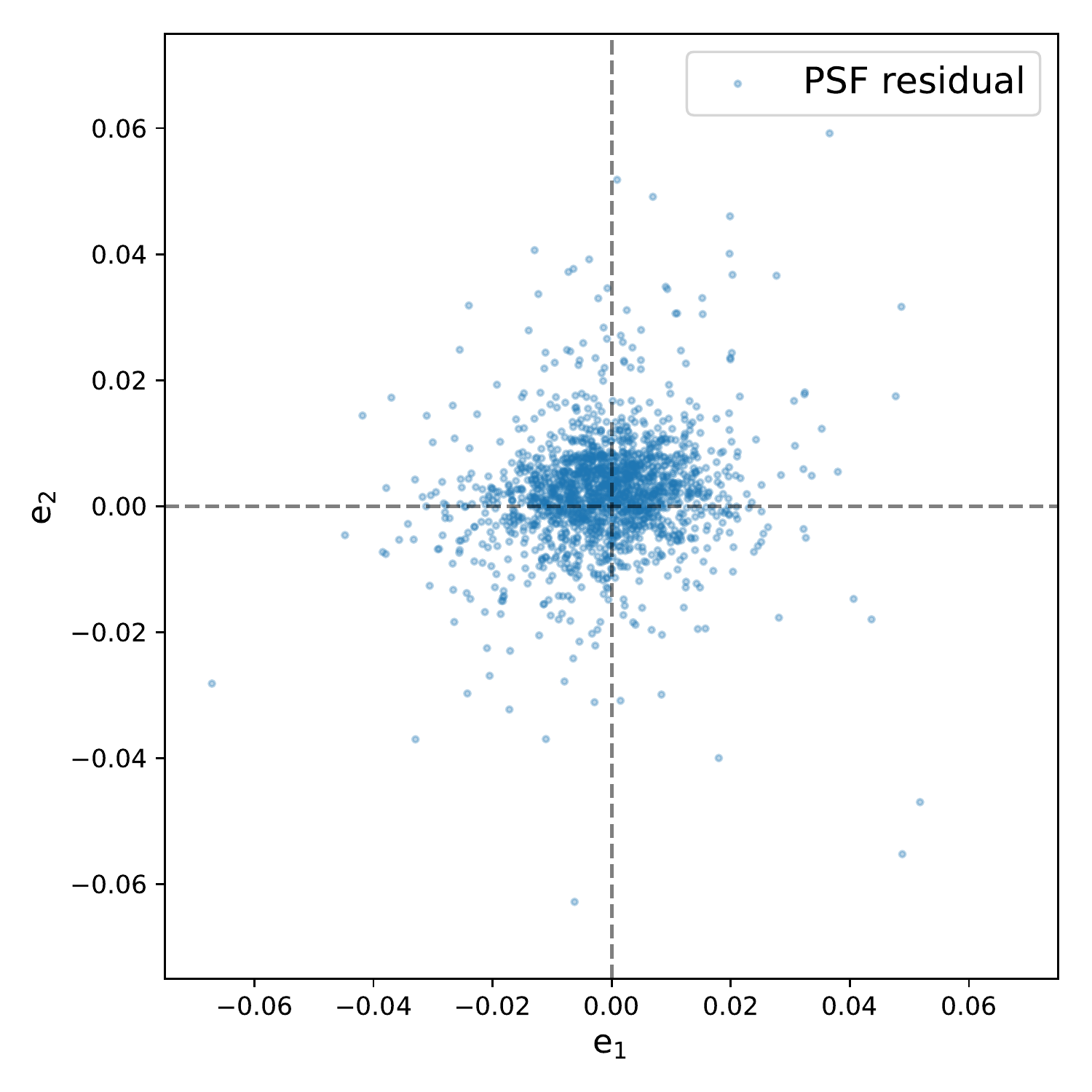}
    \caption{PSF correction in Subaru WL analysis. 
    Left: observed PSF ellipticities measured from stars in the A521 mosaic image.
    Right: residual (observed$-$model) ellipticities.
    The residual ellipticity plot shows that our PCA-based PSF model robustly represents the observed PSF ellipticities. The distribution of the residual ellipticities is much more isotropic with reduced scatters; the mean (rms) values of the $e_1$ and $e_2$ components are reduced
    from -0.014 (0.016) to -0.001 (0.010) and from 0.006 (0.009) to 0.001 (0.006), respectively.}
\label{fig:psf_correction}
\end{figure*}

Robust PSF modeling is a crucial step for accurate shape measurement. We follow the principal-component-analysis (PCA) method \citep{Jee:2007jr}, which has been extensively tested in our previous studies.
Readers are referred to Jee et al. (2007) for details of the general algorithm.
Below, we provide a description of the specific procedure
relevant to our Subaru/SuprimeCam and {\it HST}/WFC3 analyses.

When modeling the Subaru PSF in each CCD exposure frame, we use the {\tt RESAMP} images output by {\tt SWARP} \citep{2010ascl.soft10068B}.
We first selected stars based on the brightness vs. size relation. Each frame contains $\mytilde100$ high S/N stars on average, which are sufficient for modeling the spatial variation of the PSF with 3rd order polynomials discussed below.
Then, we combined the postage stamp images of the selected stars and derived the most significant 20 principal components. Each star is decomposed with these principal components and represented as a weighted sum of them. 
The weighting factor for the $i$th principal component ($C_i$) is assumed to vary spatially as follows:
\begin{equation}
   C_i(x,y) =a_{00} +a_{10}x+a_{01}y+a_{20}x^2 +a_{11}xy+a_{02}y^2 +... 
\end{equation}
\noindent
where $a_{jk}$ is the coefficient of the polynomial and we use it up to the 3rd order
($j+k \leq  3$).

After we obtained this PCA-based PSF model for individual CCDs, we created
the final ``stack" PSF model that represents the PSF in the image stack, from which we measure galaxy shapes. 
Since the {\tt RESAMP} images were already calibrated for shifts and rotations, this PSF stacking procedure is straightforward. However, it is still necessary to assign proper weights to individual PSF models according to the  contribution of each frame to the final image stack.

We display the fidelity of our Subaru PSF model in Figure~\ref{fig:psf_correction}. The residual ellipticity plot shows that our PCA-based PSF model closely represents the observed PSF ellipticities.
The distribution of the residual ellipticities is much more isotropic and
the size of the scatters are significantly reduced.

For modeling the {\it HST} PSF, the direct PCA sampling from the science image explained above is not applicable because
of the small field of view, where we have only several high S/N stellar images.
Instead, we use a template-based approach (Jee et al. 2007) to model the PSF.
First, we constructed the PSF library with PCA utilizing archival stellar field data. Then, we selected stars from the A521 images (individual exposures) based on the size-magnitude relation. Because the {\it HST} PSF pattern is related to the telescope focus breathing and thus is repeated approximately following the thermal cycle of the instrument, we find the matching template from the PSF library based on the stars on the A521 frame.
After finishing the exposure-level PSF modeling, we applied shift and rotation so that its contribution to the final mosaic image is properly accounted for. The final PSF on the mosaic image is the sum of all contributing PSFs from individual exposures.

\subsection{Source Shape Measurement}

We convolve the elliptical Gaussian galaxy profile with the model PSF and fit it to the galaxy image using the \texttt{MPFIT}\footnote{http://cars9.uchicago.edu/software/python/mpfit.html} algorithm for source shape measurement. The elliptical Gaussian model is composed of seven free parameters: centroid ($x$ and $y$), amplitude, background amplitude, semi-major and -minor axes, and orientation angle. Galaxy ellipticities ($e$) are defined as follows:
\begin{eqnarray}
 e = \frac{a-b}{a+b},\\
 e_1 = e \cos 2\theta, \\
 e_2 = e \sin 2\theta.
\end{eqnarray}
\noindent 
where $\theta$ is the orientation of the semi-major axis, and $a$ and $b$ are the semi-major and -minor axes, respectively.
The reduced shear $g_{1(2)}$ is obtained by averaging these individual ellipticities $e_{1(2)}$ as follows:
\begin{eqnarray}
g_1 = \left < e_1 \right >,\\
g_2 = \left <e_2\right >.
\end{eqnarray}

However, it has been known that the raw ellipticity measurement is a biased estimator of the true shear \citep[e.g.,][]{2015MNRAS.450.2963M}. The bias is often characterized as $g_{true} = g_{est}(1+m) + c$, where $m$ and $c$ are referred to as multiplicative and additive biases, respectively. 
Through our image simulations \citep{2013ApJ...765...74J, 2011PASP..123..596J}, 
where we created artificially lensed galaxy images matching observation properties (e.g., seeing, image depth, magnitude and size distribution, etc.) and investigated the relation between input and recovered shears, we derived a multiplicative bias of $m \simeq 0.23$  for Subaru  shapes and found that the additive bias is negligibly small ($c\lesssim 0.001$). A similar multiplicative bias $m = 0.22$ is obtained for {\it HST} data \citep{2017ApJ...847..117J}. Note that we only use the {\it HST} data for our investigation of the A521 substructure. The cluster mass measurement is solely based on the Subaru data. 

We conducted shape measurement in all three Subaru filter images ($i'$, $V$, and $R$) and verified that the resulting lensing signals are consistent. However, the final analysis is carried out with the $i'$-band image, which has the best qualities in both seeing and depth.

\subsection{Source Galaxy Selection}

\begin{table}[t]\centering
\caption{Source selection criteria}
\begin{tabular}{cc}
\hline
\hline
Magnitude       & $22 < i' \,(F140W) <27 $ \\ 
Magnitude error &  $\sigma_{i'} < 0.3$ \\ 
Magnitude error & $\sigma_V < 1 $ \\ 
Color cut        &   $ V - i' > - 1.5$ \\
(s1) Red sequence exclusion  & $\left| k_1 - 1.656 \right | > 0.35$ \\ 
(s2) Color-color cut   &   $\left| k1 - 1.656 \right | < 0.35$, $k2 > 0.2 $  \\
Ellipticity      & $ e < 0.9$ \\
Ellipticity error  & $\sigma_e < 0.25$ \\
Semi-major axis   & $a < 30$ pixel\\
Semi-minor axis   & $b > 0.3$ pixel\\
Galaxy size & $r_h>$ 1.5 pixel \\
MPFIT status &  1\\
\hline
\hline
\end{tabular}
\label{tab:source_gal_selection}
\tablecomments{See the Table \ref{tab:member_gal_selection} note for the definitions of $k_1$ and $k_2$. s1 criteria is applied to exclude the red sequence galaxies and s2 criteria is applied to reexamine the ones excluded by s1. }
\end{table}

Our source selection criteria are summarized in Table~\ref{tab:source_gal_selection}. We select source galaxies in two steps, utilizing the color-magnitude and color-color diagrams (Figure~\ref{fig:cmd_for_selection}). First, we selected faint ($22<i'<27$) galaxies whose 
$V-i'$ colors are within $0.35$ from the best-fit relation (see the dashed line in the top panel of Figure~\ref{fig:cmd_for_selection}) and excluded the red-sequence galaxies. 
Then, we examined the excluded objects using the color-color diagram and selected 
the objects whose colors do not overlap with those of the cluster member candidates (see the bottom panel of Figure~\ref{fig:cmd_for_selection}).
These sources are further required to possess well-defined ellipticities
($\sigma_e < 0.25$ and $e<0.9$) and extended shapes ($b>0.3$ pixel). 
The final source catalog
provides a density of $\mytilde 26~\mbox{arcmin}^{-2}$ for Subaru.

For the {\it HST} data, the F105W and F160W colors are not optimal to
identify and remove cluster members. Therefore, we select sources based on the F160W magnitude within the $22<\mbox{F140W}<27$ interval while removing the spectroscopic (Hectospec) and photometric (Subaru) members of A521. We also apply the same shape criteria used for the Subaru WL.
The {\it HST} WFC3-IR imaging data provide a source density of $\mytilde150~\mbox{arcmin}^{-2}$, which enables us to investigate the A521 substructure in great detail for the central $\mytilde2\farcm3 \times 2\farcm1$ region covering the first and third BCGs.

\subsection{Redshift Estimation of Source Galaxies}

A quantitative interpretation of WL signals in absolute terms requires us to define the critical surface mass density $\Sigma_c$:
\begin{equation}
    \Sigma_c = \frac{c^2}{4\pi G D_l \beta},
\label{eqn:sigma_c}
\end{equation}
\noindent
where $\beta$ is the lensing efficiency, $c$ is the speed of light, $G$ is the gravitational constant, and $D_l$ is the angular diameter distance of the cluster. The lensing efficiency $\beta$ is given by
\begin{equation}
    \beta = \bigg \langle \max \bigg(0,\frac{D_{ls}}{D_s}\bigg)\bigg \rangle. \label{eqn:beta}
\end{equation}

\noindent 
where $D_{s}$ is the angular diameter distance of sources and $D_{ls}$ is the angular diameter distance of source seen at the cluster redshift.

As we cannot obtain reliable redshift information of individual source galaxies based on three filters, we estimate the $\beta$ value for the source population by comparing our photometry with the one from a reference photometric redshift catalog. We used the Great Observatories Origins Deep Survey \citep[GOODS;][]{2004ApJ...600L..93G} photometric redshift catalogs as the reference.
By applying the same magnitude and color cuts to the GOODS catalogs, we determine $\big \langle \beta \big \rangle = 0.576$ and $\big \langle \beta^2 \big \rangle = 0.381$ from the GOODS south field and $\big \langle \beta \big \rangle = 0.577$ and $\big \langle \beta^2 \big \rangle = 0.381$ from the GOODS north field. The estimates from both fields are in an excellent agreement. The value reported in \cite{Okabe_2010} is 0.668 and 0.667 for their ``red+blue" and ``faint" objects, respectively, which implies that their sources are at higher redshift on average. 

Because in fact our sources are not located on a single redshift plane, the reduced shear is biased. 
To account for this, \cite{1997A&A...318..687S} suggested a calibration equation as the following:
\begin{equation}
g' = \left [1+\left(\frac{\left \langle \beta^2 \right \rangle}{\left \langle \beta \right \rangle^2}-1 \right)\kappa  \right]g 
\end{equation}
\noindent
where $g'$, $g$, and $\kappa$ are the uncorrected reduced shear, corrected reduced shear, and convergence, respectively. The convergence ($\kappa$) is a measure of the surface mass density in units of $\Sigma_c$.
With our measurements of $\big \langle \beta \big \rangle = 0.576$ and $\big \langle \beta^2 \big \rangle = 0.381$, the $g'/g$ ratio becomes $1+0.15 \kappa$. We apply this correction when estimating the cluster mass.

\subsection{Mass Reconstruction}
\label{sec:mass_reconstruction}
\begin{figure*}
    \centering
    \includegraphics[trim=2cm 0cm 1.2cm 1cm, width = 0.495\textwidth]{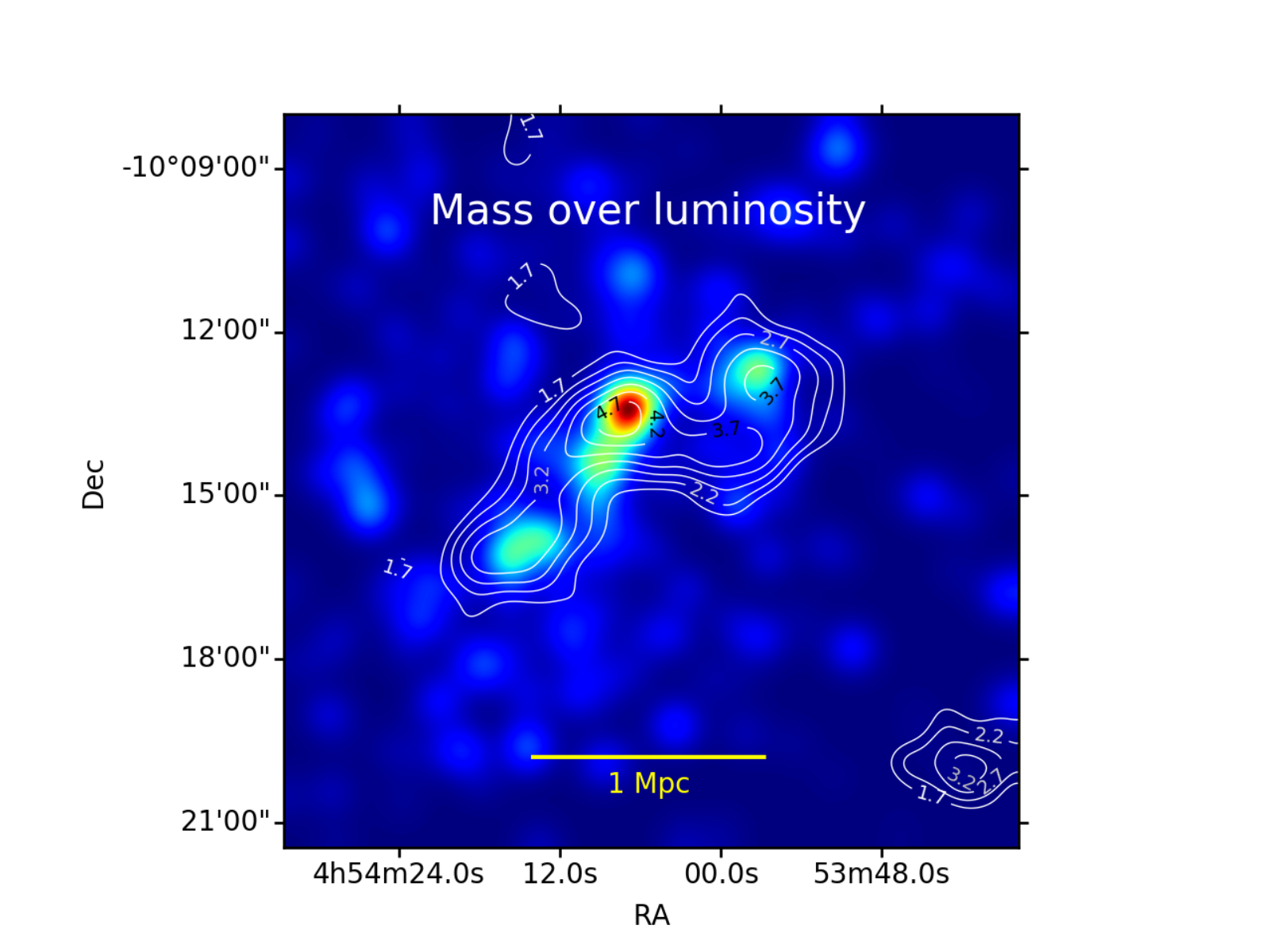}
    \includegraphics[trim=2cm 0cm 1.2cm 1cm, width = 0.495\textwidth]{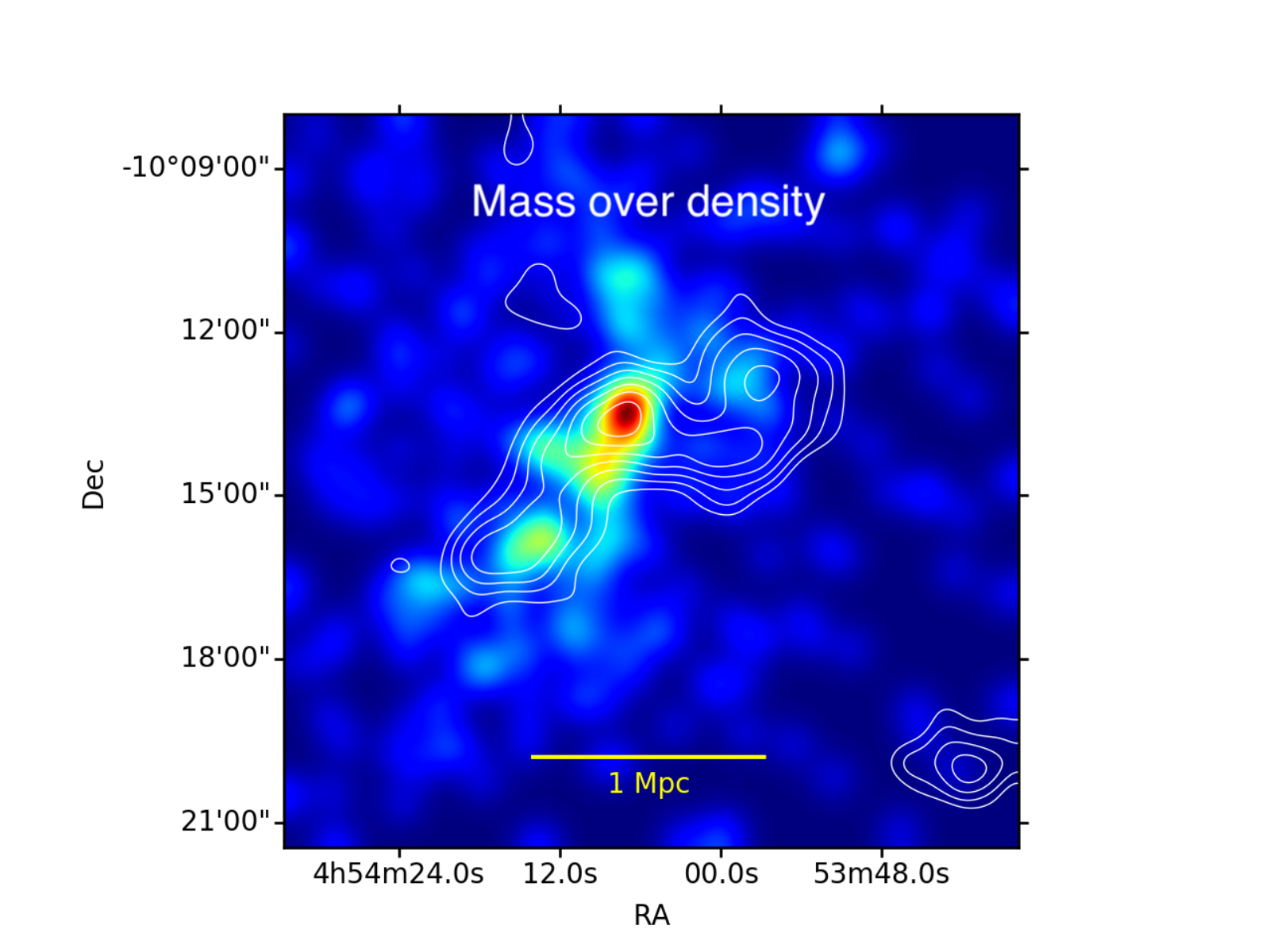}
    \caption{WL mass reconstruction of A521. 
    White contours represent the projected mass density. The effective smoothing scale is $\mytilde1.5\arcmin$. We measured the rms of the convergence to be $\sigma_{\kappa}=0.006$ from 1000 bootstrapping runs.
    The contour labels (left panel) indicate the significance ($\kappa/\sigma_{\kappa}$).
    In the left panel, color-coded is the luminosity-weighted density distribution while in the right panel, color-coded is the cluster galaxy number density.  Overall, the mass distribution agrees well with the cluster galaxy distribution. In particular, WL detects the three substructures: C, SE, and NW discussed in Figure~\ref{fig:member_galaxy_with_img}.}
    \label{fig:mass_map_with_density_map}
\end{figure*}

\begin{figure*}
\centering
 \includegraphics[trim=2.5cm 0cm 1.2cm 0cm, width = 0.48\textwidth]{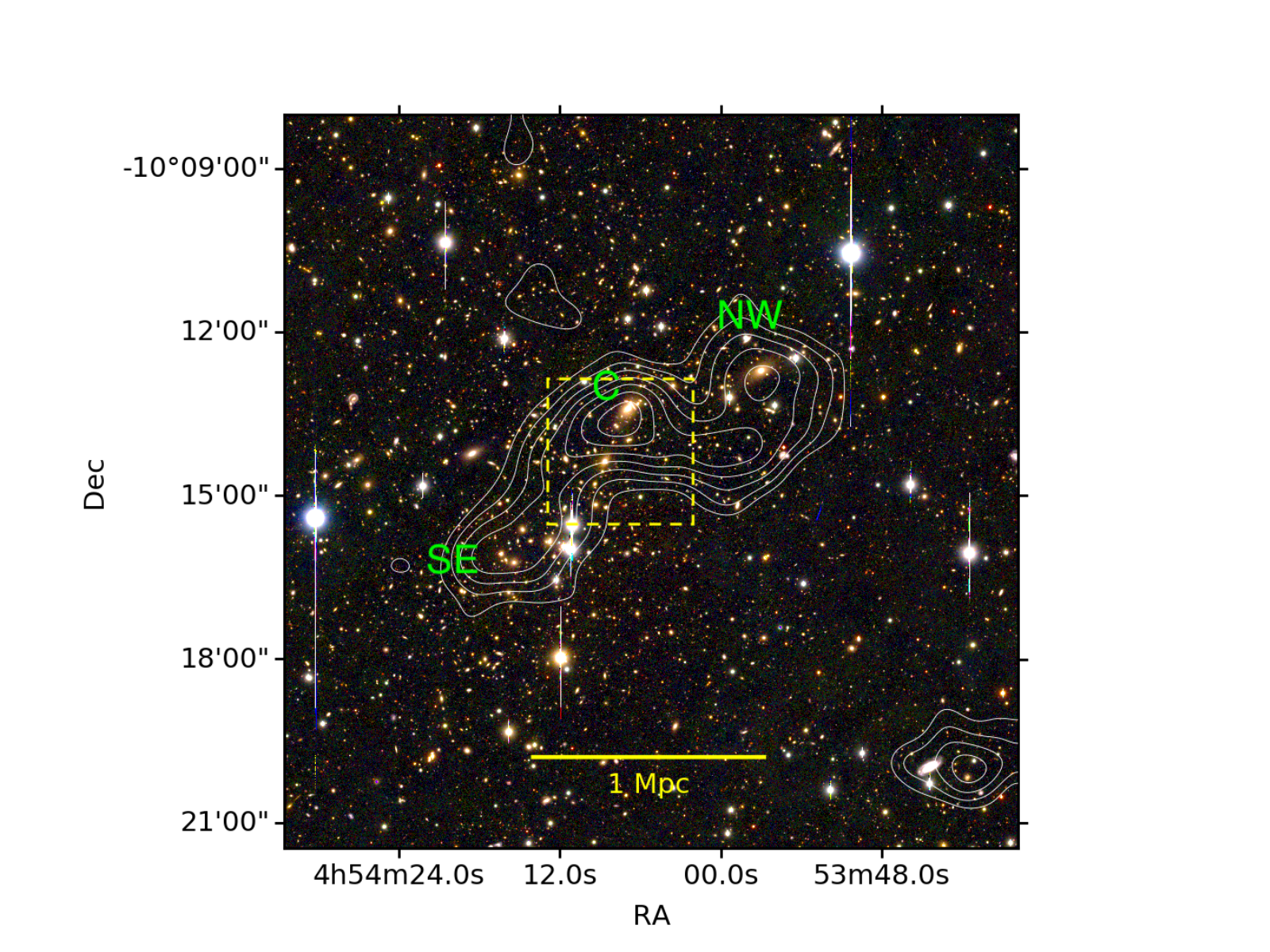}
 \includegraphics[trim=2.5cm 2.8cm 1.2cm 1cm, width = 0.45\textwidth]{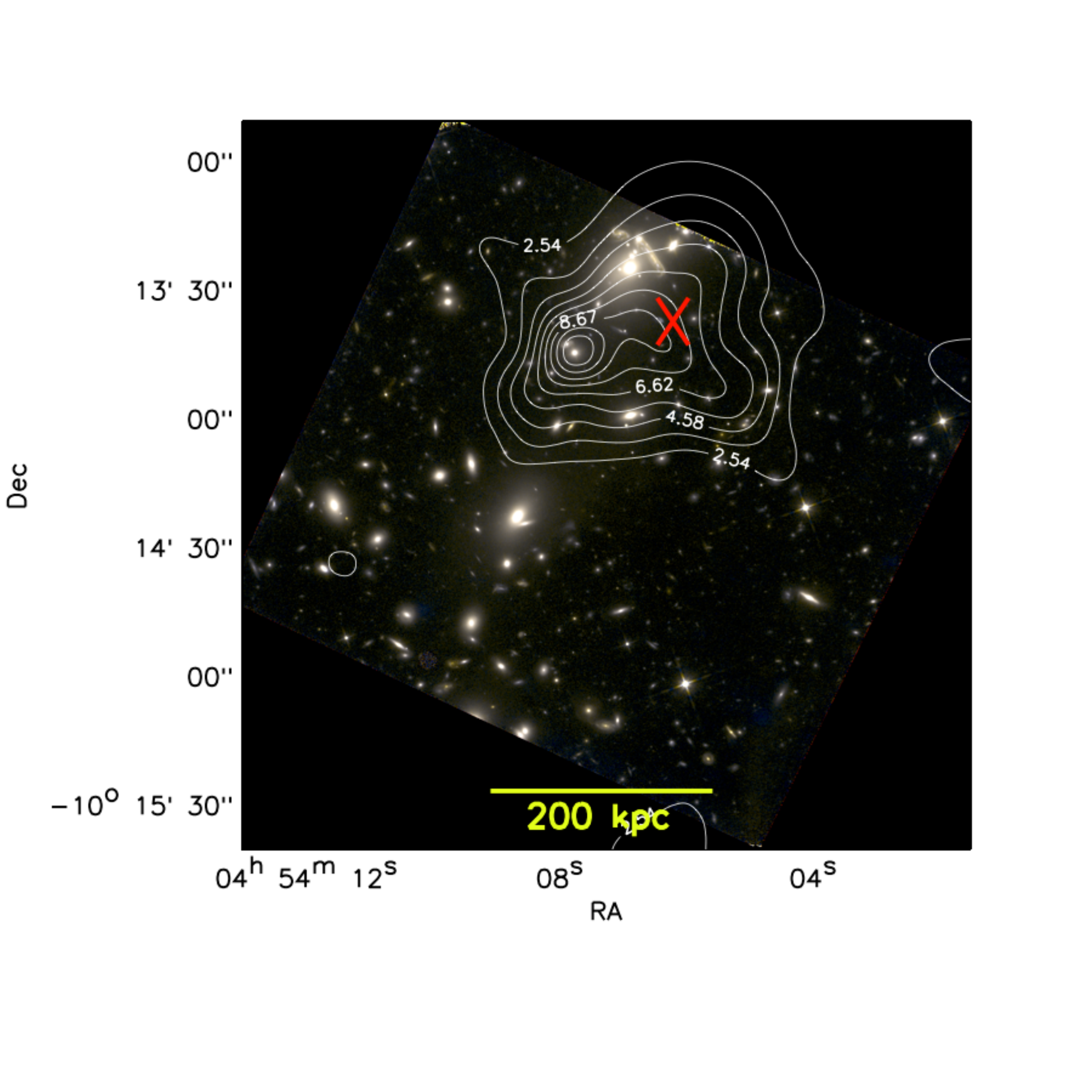}
 \caption{WL mass map overlaid on color-composite image. Left: Same as Figure~\ref{fig:member_galaxy_with_img} except that the Subaru WL mass is overlaid on the color-composite image.
 The yellow square marks the approximate region, where we perform {\it HST} 
 mass reconstruction displayed in the right panel.
 Right: We overlay the {\it HST} WL mass on the {\it HST} color-composite image. 
The contour labels indicate the significance with respect to the rms. At the location of the peak the significance is $\mytilde10~ \sigma$.
The red symbol ``X" denotes the location of the global centroid measured with the first moment using the convergence value above 3~$\sigma$. 
The HST mass reconstruction
shows that the convergence peak (where $\kappa$ is highest) is located $\mytilde25\arcsec$ east of the
global centroid. We consider the possibility that this asymmetry might arise from the superposition of two halos (one is diffuse and the other is compact).}
 \label{fig:color_map_mass_map}
\end{figure*}

The relation between convergence $\kappa$ and shear $\gamma=g(1-\kappa)$ is given by the following:
\begin{equation}
    \kappa(\mathbf{x}) = \frac{1}{\pi}\int d^2 \mathbf{x} D^*(\mathbf{x} - \mathbf{x}')\gamma(\mathbf{x}'), 
\end{equation}

\noindent 
where $D^*(x)$ is the complex conjugate of the kernal $D(x)=-1/(x_1-ix_2)^2$ when the shear $\gamma$ is expressed by the complex notation $\gamma=\gamma_1 +i \gamma_2$.
While the popular method of Kaiser \& Squires (1993) implements the algorithm in Fourier space, we use the {\tt FIATMAP} code (Fischer \& Tyson 1997), which performs the computation in real space. The method uses a modified kernel, which minimizes the edge effect and addresses the nonlinearity $g=\gamma/(1-\kappa)$.
The reconstructed mass distribution is overlaid on galaxy distributions in Figures~\ref{fig:mass_map_with_density_map} and~\ref{fig:color_map_mass_map}. The effective smoothing scale is $\mytilde1.5\arcmin$. With bootstrapping analysis, we estimated the rms of the convergence is $\sigma_{\kappa}\simeq 0.006$. The contour labels in Figures~\ref{fig:mass_map_with_density_map} and~\ref{fig:color_map_mass_map} represent the significance ($\kappa/\sigma_{\kappa}$).

Remarkably, the reconstructed mass map closely follows the galaxy distribution, detecting the three substructures discussed in \textsection\ref{sec:phot_cluster_member}. 
Our booststrapping analysis shows that the mass and luminosity centroids agree within 1 $\sigma$ (Table~\ref{Tab:sub_clump_mass}).
The two WL centroids of the C and NW clumps coincide with those of the brightest cluster galaxies (see left panel of Figure~\ref{fig:color_map_mass_map}). We cannot identify a distinctively bright galaxy representing the centroid of the SE clump. Nevertheless, its luminosity centroid is also in good agreement with its WL mass peak (see left panel of Figure~\ref{fig:mass_map_with_density_map}). 

\cite{Okabe_2010} also conduct WL analysis on A521. Although they present a mass map showing the substructures, no discussion pertaining to the features
is given in the paper. In their mass map, two dominant mass clumps are seen. One main clump coincides with our C clump. The other clump is centered at the star, $\mytilde1\farcm3$ west of our SE peak. Our NW clump does not have a corresponding feature in their map. 

The right panel of Figure \ref{fig:color_map_mass_map} shows the {\it HST} WL mass reconstruction result. Because it has much higher source density $\mytilde150$~arcmin$^{-2}$ than the Subaru value $\mytilde26$~arcmin$^{-2}$, the significance is also much higher ($\mytilde10~ \sigma$ at the peak for {\it HST} while $\mytilde5~\sigma$ for Subaru). The result shows that only a single mass peak is present near the C clump in-between the the first and third BCGs as seen in the Subaru result. Note that the peak (where the convergence value is highest) is located $\mytilde25\arcsec$ east of the large-scale centroid (marked with the symbol ``X"), which
was measured from the first moment of the significant ($>3~\sigma$) $\kappa$ region.
We discuss this feature in the context of the merging scenario reconstruction in \textsection\ref{sec:mergining_scenario}.

\subsection{Mass Estimation}

\subsubsection{Mass Estimation from Tangential Shear}
\label{sec:tangential_shear_mass_est}
\begin{figure}
    \centering
    \includegraphics[width = 0.49\textwidth]{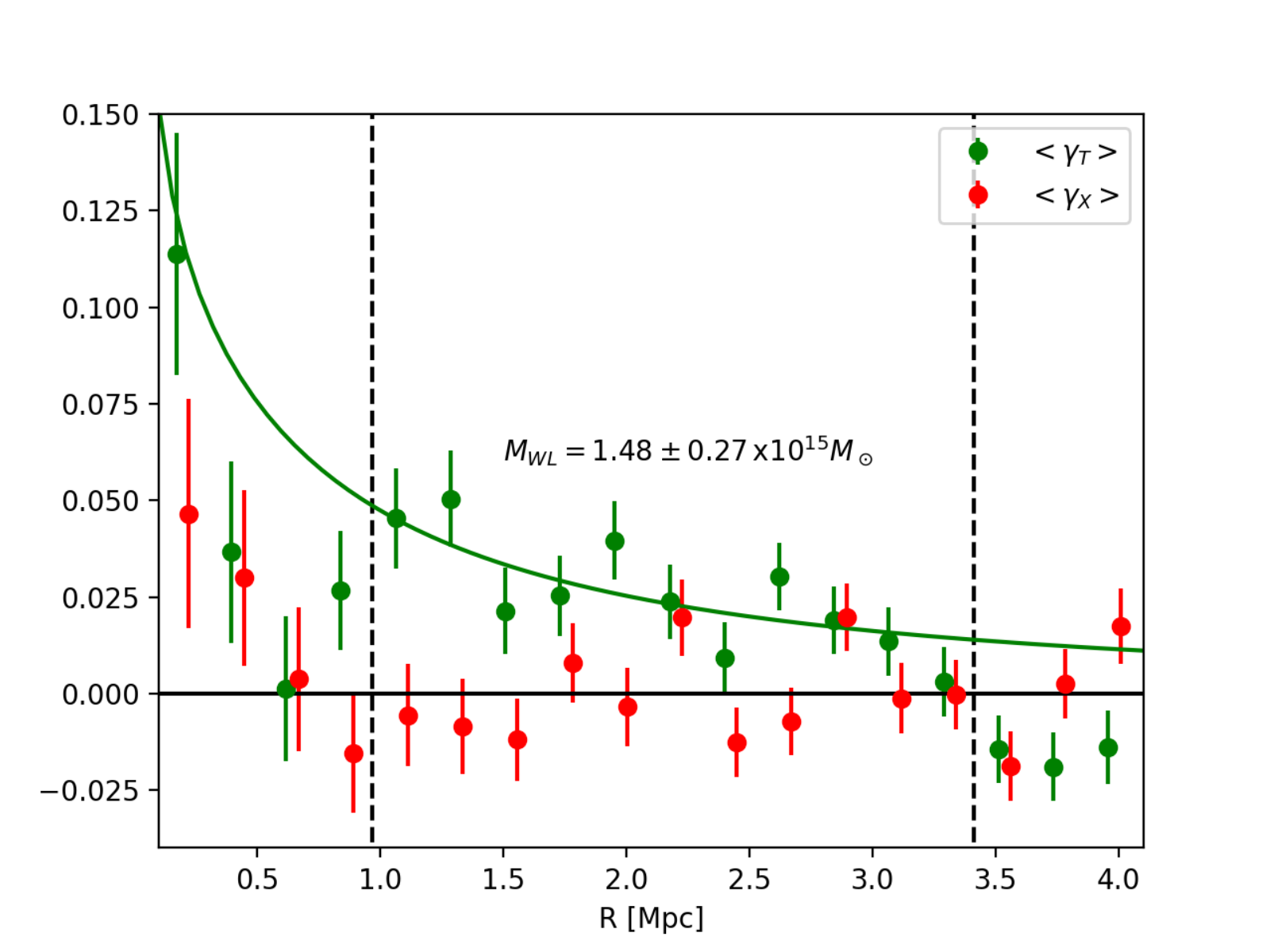}
    \caption{Tangential shear (green) and cross shear (red) radial measurement around the cluster center (BCG 1). The cross shear is consistent with zero at most radio bins except for the bin around 3~Mpc. A single NFW profile is fitted to the tangential shear profile ranging from 1~Mpc to 3.3~Mpc. The best-fit mass is $M_{200}=1.48\pm 0.27 \times 10^{15} M_\odot$.}
    \label{fig:NFW_fitted_shear}
\end{figure}

Under the assumption that a cluster consists of a single halo, one can estimate the total mass of the cluster by fitting an analytic profile to the tangential shear measurement. Although we find that A521 is comprised of multiple halos, we include the cluster mass obtained in this way in our presentation of the results in order to enable comparisons with previous studies. However, we remind the reader that our main results are the multi-halo fitting results described in \textsection\ref{sec:multi_halo_fitting}.

As shown in the Figure~\ref{fig:NFW_fitted_shear}, we fit an Navarro-Frenk-White profile \citep[NFW;][]{1997ApJ...490..493N} to the tangential shear measurements around the BCG. We only use the measurements from 1~Mpc to 3.4~Mpc. The inner cutoff radius is set to minimize the contamination due to the substructures NW and SE (see Figure \ref{fig:color_map_mass_map}). The choice of the outer cutoff radius is somewhat arbitrary since the current Subaru field is much larger than the cluster virial radius. Our choice of $\mytilde3.4~$Mpc roughly corresponds to $\mytilde1.5$~$r_{200}$.
Using the mass-concentration relation from \cite{2008MNRAS.390L..64D}, we estimate the total mass of the cluster to be $M_{200} = 1.48 \pm 0.27 \times 10^{15} M_{\odot}$. 
This mass estimate is more than a factor of two higher than the one from \cite{Okabe_2010}, who quoted $M_{200} = 6.54^{+1.43}_{-1.26} \times 10^{14} M_{\odot}$\footnote{We used $h = 0.7$ to convert the $M_{200}$ value in Table 8 of \cite{Okabe_2010}.
}. 

\subsubsection{Mass Estimation from Multi-halo Fitting}
\label{sec:multi_halo_fitting}

In \textsection\ref{sec:tangential_shear_mass_est} we present a rough estimate of the A521 mass by treating the cluster as a single halo. Here, we provide a more accurate analysis of the cluster mass by fitting multi-halo models. We measure the posterior using the nested sampling algorithm \citep{2009MNRAS.398.1601F}.

As we identify three clumps in the reconstructed mass map, whose centroids are in excellent agreement with the luminosity peaks (Figure~\ref{fig:mass_map_with_density_map}), we simultaneously fit three NFW halos with their centroids fixed at the luminosity peaks. Since each NFW profile is determined by two parameters (when its cetroid is fixed), six parameters are needed in total. We choose mass and concentration to represent each NFW halo. For the analysis of the likelihood sampling, our mass and concentration priors are set to the ranges $10^{13} M_{\sun} \leq M_{200} \leq 10^{15} M_{\sun}$ and $ 2 \leq c \leq 5$, respectively. The mass estimates ($M_{200}$) of the C, NW, and SE clumps are $4.38^{+0.92}_{-0.87}\times10^{14}M_{\sun}$, $2.37^{+0.68}_{-0.46}\times 10^{14}M_{\sun}$, and $2.56^{+0.90}_{-0.51}\times 10^{14} M_{\sun}$, respectively. Assuming the three clumps are located at the same distance from us, we can superimpose the three halos' density profiles to estimate the total mass. We find that the $r_{200}$ radius (inside which the mean density of the sphere is 200 times the critical density at the cluster redshift) is $r_{200}=2.07^{+0.06}_{-0.07}~$Mpc. The resulting mass is $M_{200}=13.0^{+1.0}_{-1.3}\times~10^{14}~M_{\sun}$. We did not achieve meaningful constraints on concentration parameters.

Traditionally, mass-concentration relations \citep[e.g.,][]{2008MNRAS.390L..64D,2014MNRAS.441.3359D} have been employed to reduce the number of free parameters when NFW halos are used. The relation represents a mean behavior for a population of clusters with large scatters in numerical simulations and thus may not be applicable to individual clusters. When we select the \cite{2008MNRAS.390L..64D} relation,
the mass estimates of the C, NW, and SE clumps are $M_{200}=5.06^{+1.21}_{-0.80}\times 10^{14}M_{\sun}$, 
$2.23^{+0.78}_{-0.50} \times 10^{14}M_{\sun}$,
and $2.28^{+0.90}_{-0.44} \times 10^{14}M_{\sun}$, respectively, all of which are consistent with the ones derived without the mass-concentration relation.
We summarize the mass estimates above in Table~\ref{Tab:sub_clump_mass} and \ref{Tab:Total_mass}.

\begin{table*}\centering
\caption{Mass estimation and velocity of substructures}
\scriptsize
\begin{tabular}{cccccccccc}
\hline
\hline
Clump & spec-$z$  & RA$^{1}$& Dec    &  mass centroid$^2$ & $\Delta v^3$ &  $\sigma_{v}^4$  & $M_{200}$ (w/o $M-c$) & $M_{200}$ (w/ $M-c$)&  peak \\
 &number &&& uncertainty&[$\kms$] & [$\kms$]& [$10^{14} M_{\odot}$] & [$10^{14} M_{\odot}$] &significance\\
\hline
C & 30 &  $4^h54^m8.3^s$& -$10\degr13\arcmin46.5\arcsec$  & 20.43$\arcsec$ & 14 $\pm{258}$ & 1390$\pm{183}$  & $5.06^{+1.21}_{-0.80}$&$4.38^{+0.92}_{-0.87}$ & 5.2 $\sigma$\\
 & & ($4^h54^m6.9^s$)& (-$10\degr13\arcmin26.2\arcsec$)  & &  &   &  & \\
NW & 12 &   $4^h53^m57.6^s$& -$10\degr12\arcmin57.9\arcsec$ &27.5$\arcsec$ & -588$\pm{379}$ &  1256$\pm{268}$  &  $2.23^{+0.78}_{-0.50}$ &$2.37^{+0.68}_{-0.46}$ & 3.7 $\sigma$\\
 & & ($4^h53^m57.3^s$)& (-$10\degr12\arcmin42.6\arcsec$ ) & &  &   &  & \\
SE & 10 &  $4^h54^m16.1^s$& -$10\degr16\arcmin12.2\arcsec$ & 28.8$\arcsec$ & -183 $\pm{265}$& 794 $\pm{197}$& $2.28^{+0.90}_{-0.44}$ & $2.56^{+0.90}_{-0.51}$ & 3.5 $\sigma$\\
 & &  ( $4^h54^m14.7^s$)& (-$10\degr15\arcmin56.4\arcsec$)  & &  &   &  & \\
\hline
\hline
\end{tabular}
\tablecomments{$^1$ The coordinate of the mass (luminosity) peak is shown without (with) parenthesis.
The luminosity peaks agree with the mass peaks at the 1$\sigma$ level.
$^2$ We determine the mass centroid uncertainty from 1000 bootstrapping runs. 
$^3$ The reference is the mean redshift of A521.
$^4$ The LOS velocity dispersion is based on spectroscopic members.
}
\label{Tab:sub_clump_mass}
\end{table*}

\begin{table}\centering
\caption{Total mass estimation of A521 }
\begin{tabular}{cc}
\hline
\hline
Halo modeling & $M_{200}$ [$10^{14} M_{\odot}$] \\ 
\hline

One halo with $M-c$ relation  & 14.8 $\pm$ 2.7 \\
Multi halo with $M-c$ relation  & 10.4$^{+0.6}_{-0.7}$ \\
Multi halo without $M-c$ relation & $13.0^{+1.0}_{-1.3}$ \\

\hline
\hline
\end{tabular}
\label{Tab:Total_mass}
\end{table}


\section{Discussion}
\label{sec:discussions}

\subsection{Mass Comparison with Previous Studies}

We determine the total mass of A521 to be $M_{200}=1.30^{+0.10}_{-0.13}\times 10^{15} M_{\sun}$ by modeling the system with three halos. 
The error bars here only include statistical uncertainties. Considering the various systematic uncertainties discussed in \cite{2014ApJ...785...20J} such as the miscentering, triaxiality, large-scale structure, departure from NFW profiles, etc., we estimate that the total
mass uncertainty of A521 would increase up to $\mytilde20$\% of the total mass.
Nevertheless, our WL mass is significantly (a factor of two) larger than the WL mass estimate $M_{200} = 6.54^{+1.43}_{-1.23} \times 10^{14}  M_{\odot}$ from \cite{Okabe_2010}. In their analysis, they treat the A521 system as a single halo and use tangential shear measurements without excluding the shear signal at small radii ($\lesssim$ 1~Mpc). Because of the complicated substructure, the shears in the inner region are suppressed. Thus, using these suppressed signals while assuming a single halo can lower the mass estimate. In our case, where we exclude the signal at the inner radii, our one-halo mass estimate is consistent with the one from the multi-halo modeling.

Our WL mass estimate is $\mytilde 34$\% lower than the dynamical mass estimate $M_{vir} = 1.96\times 10^{15} M_{\odot}$ of \cite{2003A&A...399..813F}. However, since the authors do not quote an uncertainty of their mass estimate, we cannot evaluate the significance of the difference. 
The Planck SZ-based mass estimate \citep{2015A&A...581A..14P} is
$M_{yz, 500} = 6.90_{-0.61}^{+0.64}\times 10^{14} M_{\odot}$, which is $\mytilde23$ \% lower than our result
\footnote{Our conversion to $M_{500}$ yields $M_{500}=8.92^{+0.69}_{-0.89}\times10^{14}M_{\sun}$.}.
Since A521 departs from the dynamical equilibrium and consists of multiple clumps, these mass estimates based on the velocity dispersion and SZ measurements are potentially biased.

\subsection{Substructures}

\begin{figure*}[ht]
    \centering
    \includegraphics[trim=3cm 0cm 0cm 0cm,width = 0.40\textwidth]{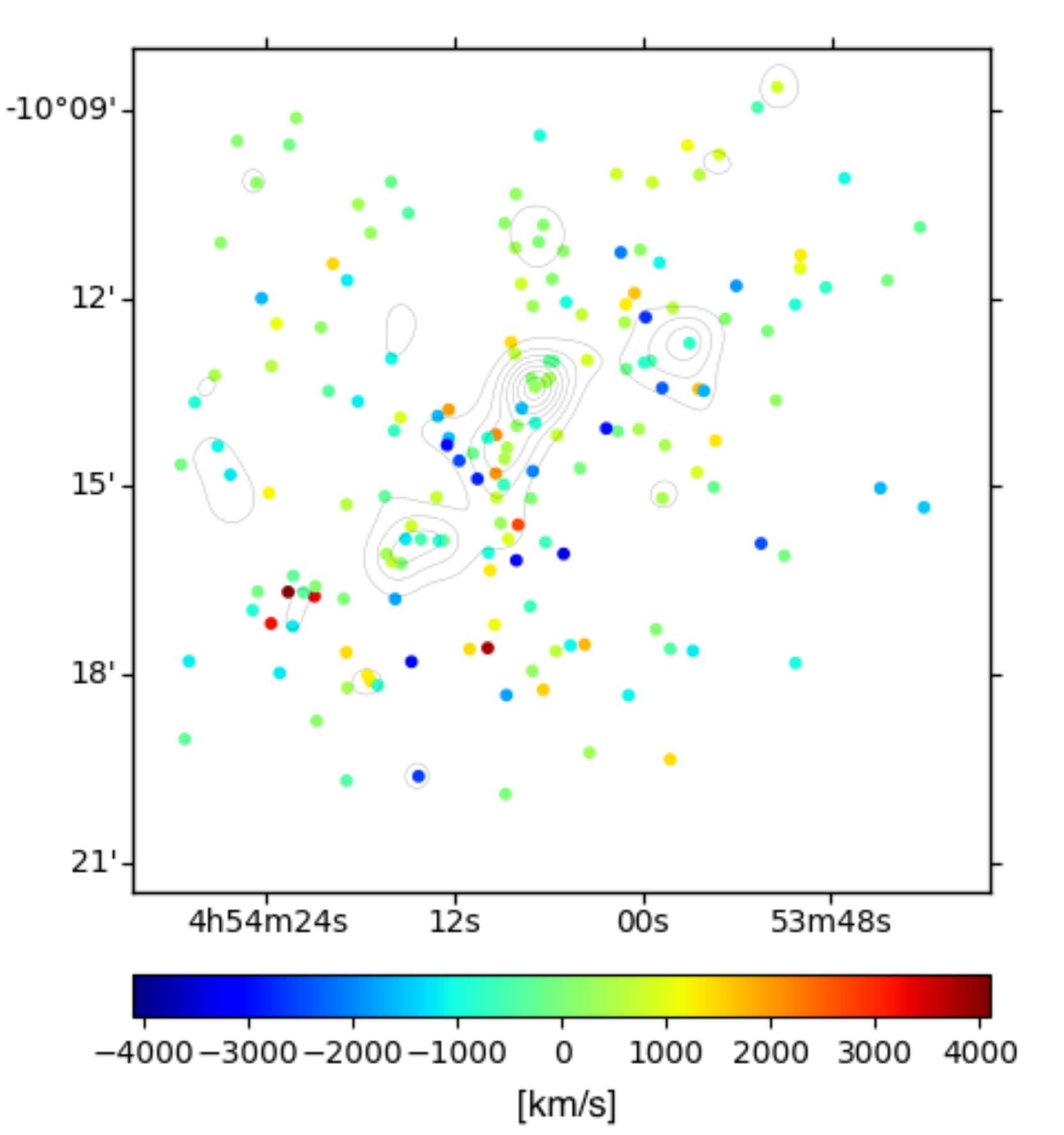}
    \includegraphics[trim=0.5cm -0.78cm 2cm 1cm, width = 0.398\textwidth]{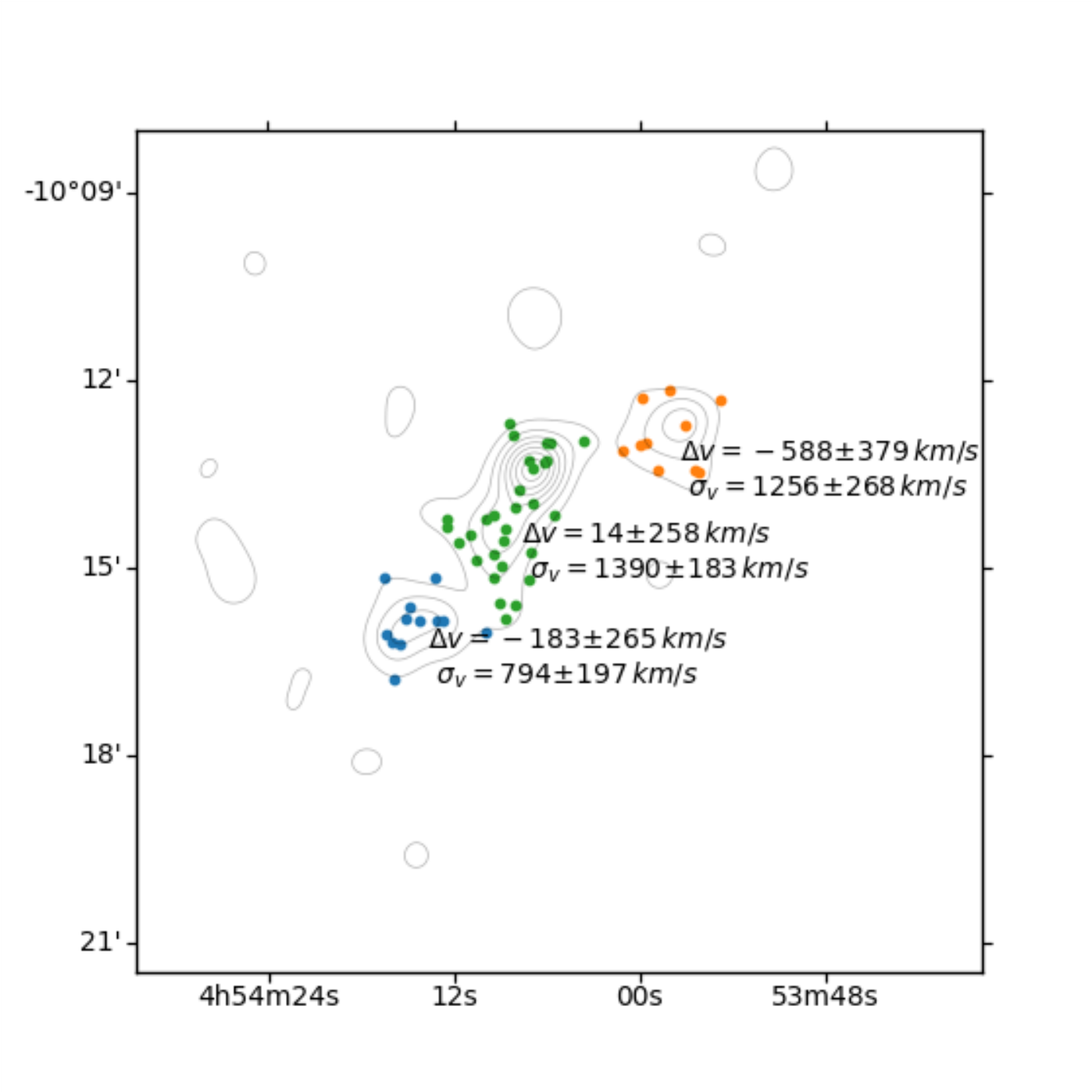}
    \caption{Luminosity-weighted galaxy contours and spectroscopic members belonging to the NW, C, SE clumps. The contours are linearly spaced with the lowest level corresponding to 15\% of the maximum.
    Left: the color represents relative velocity of each spectroscopic member with respect to the mean cluster velocity. Right: the NW, C, SE clump members are depicted in orange, green, and blue respectively with each clump's mean relative velocity ($\Delta  v$) and velocity dispersion ($ \sigma_{v}$). No statistically significant LOS velocity difference is found among the three galaxy groups. }
    \label{fig:sub_group_vel}
\end{figure*}

\begin{figure*}[h]
    \centering
    \includegraphics[trim=2cm 0cm 1.2cm 1cm, width = .495\textwidth]{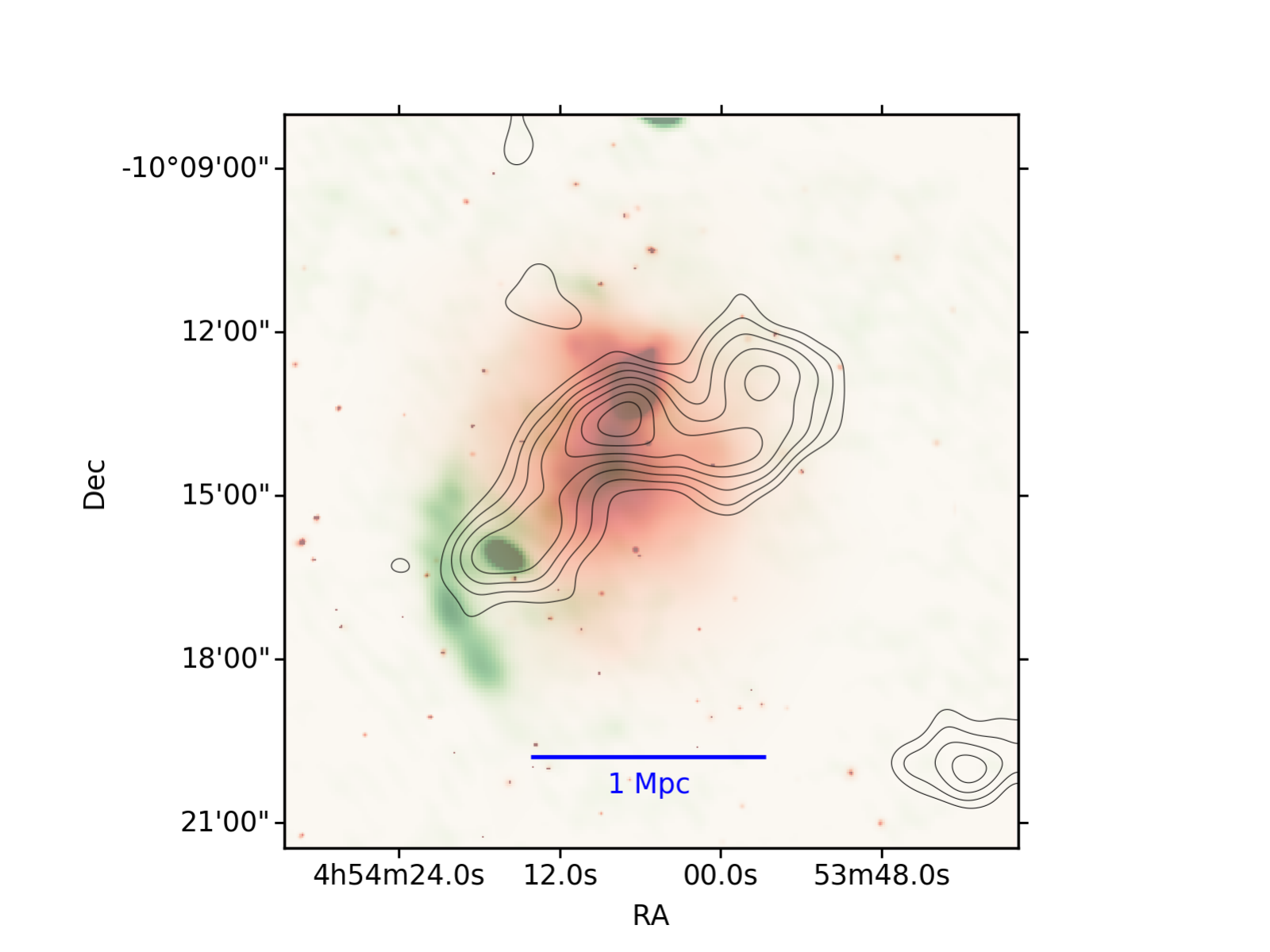}
    \includegraphics[trim=2cm 0cm 1.2cm 1cm, width = .495\textwidth]{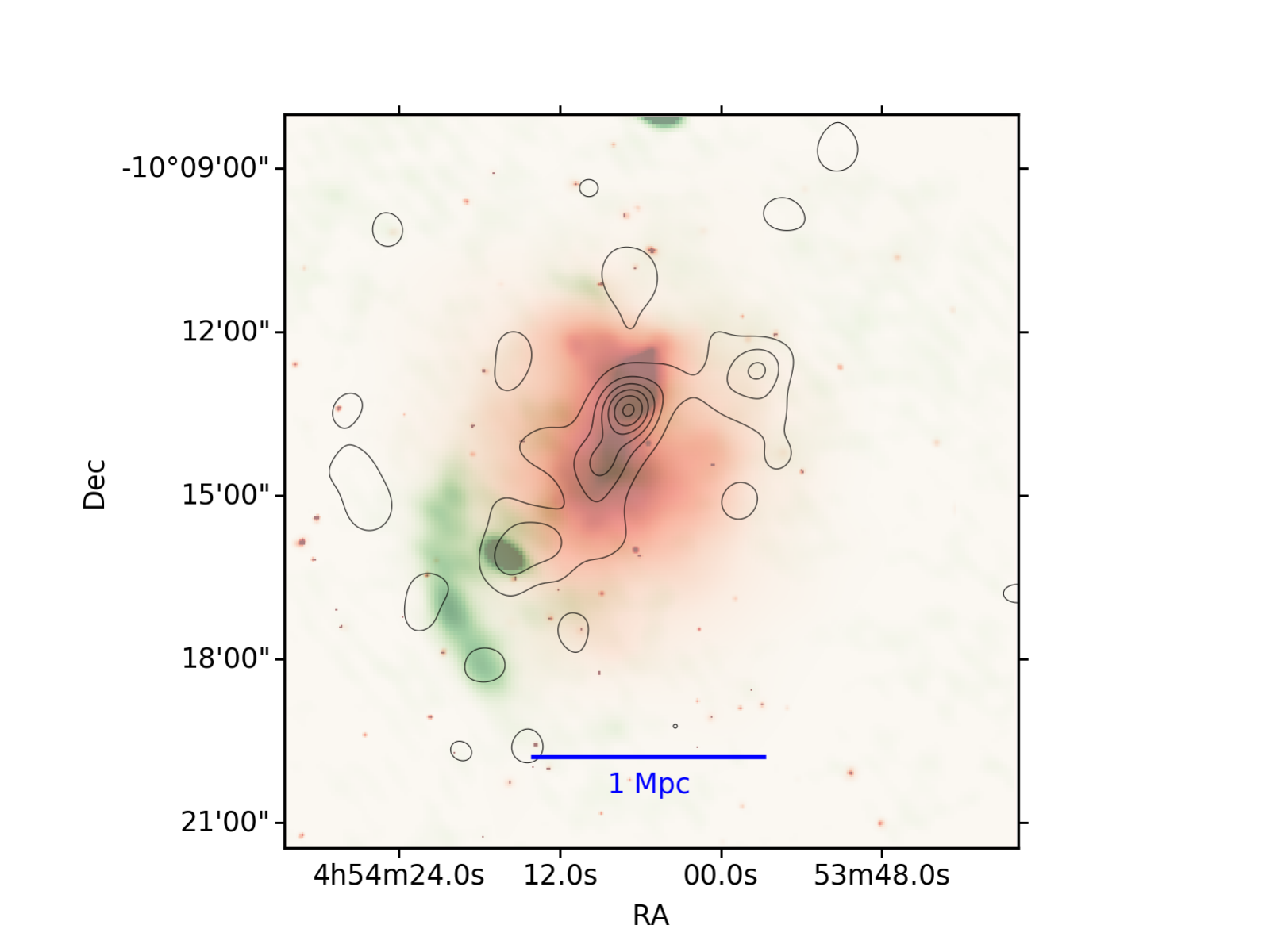}
    \caption{GMRT radio at 153MHz (green) and {\it Chandra} X-ray (red) images overlaid with WL mass map contours (left) and luminosity contours (right). In the central region, while WL detects only one mass clump (the C clump), X-ray observations reveals two peaks bracketing the mass clump. The optical luminosity distribution hints at this bimodality seen in X-ray albeit less prominent. 
    The location of the radio relic is \mytilde 1 Mpc away from the cluster center. The NW and SE mass/galaxy clumps are not associated with any significant X-ray emission.
    Based on these multi-wavelength data, we propose two merging scenarios (see text).
    }
    \label{fig:radio_xray_mass_map}
\end{figure*}

As mentioned in \textsection\ref{sec:mass_reconstruction}, the peak locations in the luminosity-weighted galaxy distribution are in excellent agreement with the WL mass centroids (Figure~\ref{fig:mass_map_with_density_map}). Our scrutiny reveals that a noteworthy difference in morphology is found for the C clump.
While both centroids agree, the luminosity clump has a tail extended toward southeast as opposed to the mass clump showing an extension toward west. 
This mass extension lacks optical counterparts. The WL substructure remains significant in our experiment with bootstrapping and different source selection, although it is difficult to interpret the feature.
When we apply a smaller smoothing kernel to the luminosity map, the C clump is further resolved into two smaller clumps with the larger of the two centered on the first BCG and the other on the third BCG  ($\mytilde300$~kpc southeast of the first BCG). This bimodality is consistent with the X-ray image that also shows the northern and southern peaks in the C clump region. (see Figure~\ref{fig:radio_xray_mass_map}). 
Our Subaru WL detected only a single mass clump between the two X-ray peaks, which is confirmed by our {\it HST} analysis. The {\it HST} WL resolution ($\mytilde40\arcsec$) is capable of resolving two $\mytilde10^{14}~M_{\sun}$ mass clumps if they are separated by $\mytilde300$~kpc ($\mytilde1.3\arcmin$) \citep[e.g.,][]{2005ApJ...618...46J, 2005ApJ...634..813J}.

We consider two possibilities to resolve this puzzle.
One possibility is that there exist two mass clumps whose separation is much smaller than our WL resolution. 
The other is that the C clump is associated with the northern X-ray peak while the southern X-ray peak is linked to the SE clump. For these two cases, we discuss the corresponding merging scenarios using our numerical simulations in \textsection\ref{sec:mergining_scenario}.

We measured the mean velocity and velocity dispersion of the cluster galaxies belonging to the three substructures. Spectroscopic members of each clump were determined with the boundary defined by the lowest-level iso-luminosity contour in Figure~\ref{fig:sub_group_vel} (corresponding to the 15\% of the maximum value).
The resulting object numbers for the C, NW, and SE clumps are 30, 12, and 10, respectively.
The velocity dispersion of the C, NW, and SE clumps are 1390$\pm{183}$, 1256$\pm{268}$, and 794$\pm{197}$~\kms, respectively. The velocity dispersions of the NW and SE substructures are difficult to interpret because of the small numbers of the used spectroscopic members. The C clump has a high velocity dispersion, which corresponds to an extremely high dynamical mass estimate $\gtrsim 3.9 \times 10^{15} M_{\sun}$ \citep{Saro_2013}. This value is much higher than our WL mass estimate $\mytilde5\times10^{14}~M_{\sun}$. One interpretation of this large discrepancy is that the velocity dispersion is inflated because of the merger. \cite{1996ApJS..104....1P} showed that when a 3:1 merger is viewed perpendicular to the merger axis, the post-merger LOS velocity dispersion $\sigma_{post}$ can increase by up to $\mytilde30$\%, compared to the pre-merger value $\sigma_{pre}$. Some observations report similar values. For example, \cite{2017MNRAS.466.2614M} report boost factors ($f=\sigma_{post}/\sigma_{pre}$) of $1.26^{+0.18}_{-0.19}$ and $1.18_{-0.20}^{+0.23}$ for the NW and NE systems of A1758, respectively. However, the boost factor for the C clump in A521 is somewhat extreme ($f\sim1.9$).
A similarly large difference is reported in \cite{Kim_2019}, who studied A115 with multi-wavelength data. They find that the WL masses are an order of magnitude lower than what the velocity dispersions imply, attributing the large differences to the on-going merger activities.

Table~\ref{Tab:sub_clump_mass} lists the relative LOS velocity $\Delta v$ of each clump with respect to the mean velocity of the entire system.
The relative velocities of the three clumps are consistent with our hypothesis that in general cluster mergers identified with radio relics are happening in the plane of the sky with small LOS velocities. However, note that small relative velocities alone do not exclusively mean the plane-of-the-sky merger because any relative velocity (even if it is very large) along the LOS direction eventually vanishes at the epoch when the subcluster reaches the apocenter \citep{2016ApJ...831..110G}.

\begin{figure*}[h]
    \centering
    \includegraphics[width = 1.\textwidth]{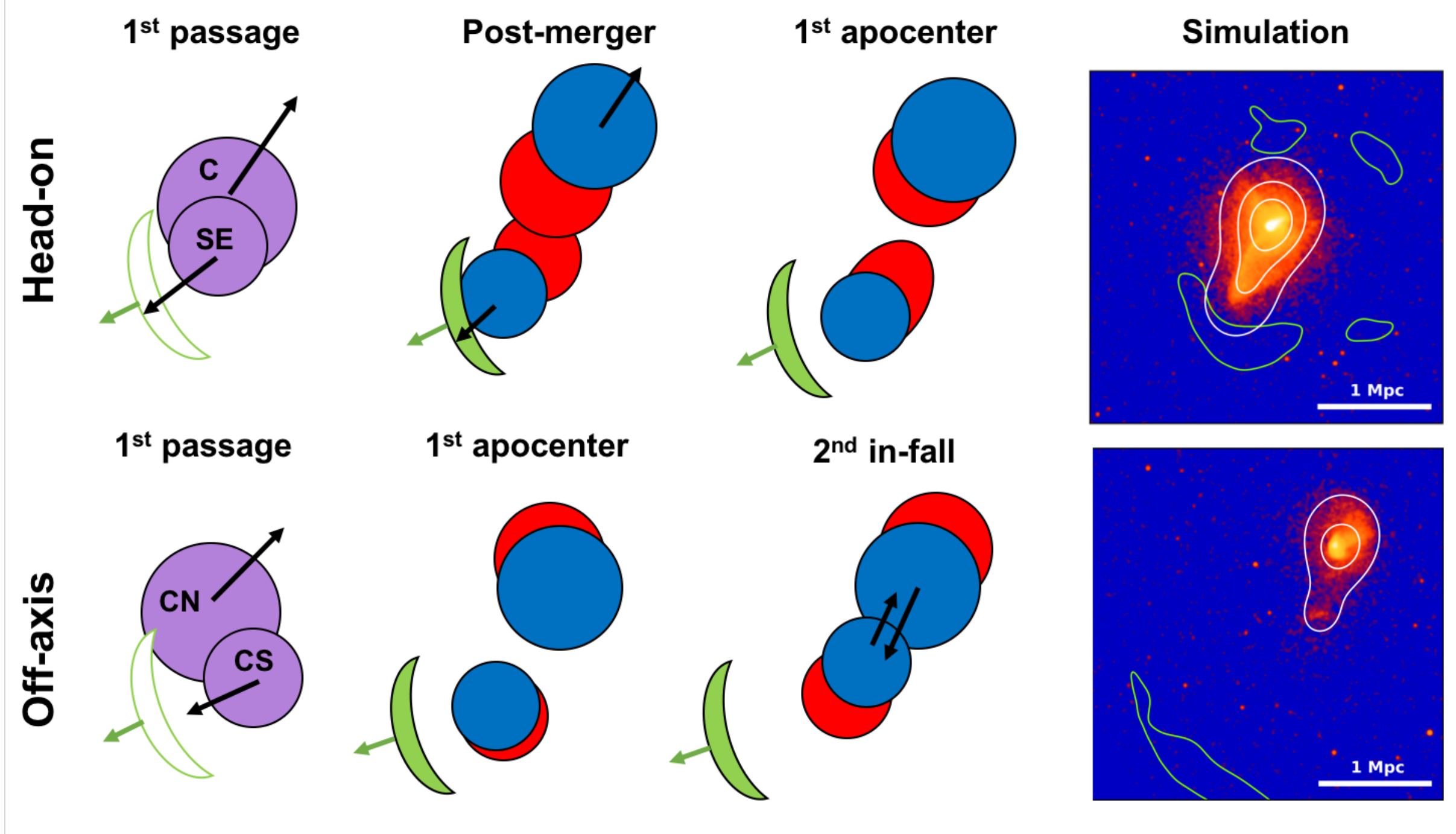}
    \caption{Schematic diagrams and toy simulation results of our two A521 merging scenarios.
    The first column illustrates the substructure configuration of the A521 dark matter and gas at the first passage.  Purple circles represent the halos without dark matter-ICM dissociation.
    In the second and third columns, blue and red circles represent the dark matter and ICM, respectively, while  green arcs indicate the merger shocks (radio relics). In the last column, we display the mock {\it Chandra} ACIS-I X-ray map (red) of a cluster at $z=0.25$ generated by {\tt pyXSIM} with the assumption of 100 ks exposure.
    White contours indicate the projected mass (dark matter plus ICM) with linear spacing. Green contours marks the locations of the hot outskirt region ($ T > 10\rm\,keV, r > 0.5\rm\,Mpc$) in the collision plane. The X-ray, mass, and temperature maps are smoothed with the kernel size of $11 \rm\,kpc$, $120\rm\,kpc$, and $60\rm\,kpc$, respectively. Top row describes the case for the head-on collision between the C and SE clumps, where the simulated A521 analog appears at the first apocenter. Bottom row shows the off-axis collision between
    two halos, which comprise the C clump. We depict the northern and southern parts as CN and CS, respectively.
    In this latter case, the simulated A521 analog happens at the second collision.}
    \label{fig:revised_scenario}
\end{figure*}

\subsection{Revision of Merging Scenario with Numerical Simulations}
\label{sec:mergining_scenario}
One of the crucial prerequisites for a merging scenario reconstruction is identification of the subclusters involved in the merger. The most common method is to examine the difference between the distributions in the ICM and cluster galaxies. This approach, however, should be used with caution in A521, which show non-trivial complexity in both cluster constituents. 

\cite{2006A&A...446..417F} interpreted the  Chandra X-ray data as indicating at least two merger events: a pre-merger of the southern main and northern infalling components and a post-merger along the NE-SW ridge (see Figure~\ref{fig:member_galaxy_with_img} for the location and orientation of the ridge). The post-merger argument is based on the high velocity dispersion and temperature structure of the region whereas the pre-merger hypothesis is based on the X-ray tail morphology of the northern component and the gas compression feature at the southern edge of the BCG.

It is difficult to reconcile the pre-merger scenario of \cite{2006A&A...446..417F} with the position and location of the radio relic. The reality of the radio relic is confirmed by \cite{2008A&A...486..347G}, who with GMRT and VLA observations find that the radio relic has a spectral steepening toward the cluster, which indicates the shock propagation (thus the merger) direction. 
\cite{2008A&A...486..347G} also report that the {\it Chandra} data show a surface brightness jump across the radio relic.
Based on XMM-{\it Newton} data, \cite{Bourdin_2013} find  two X-ray shock features. One coincides with the location of the radio relic and the other is located at the southwestern edge of the X-ray emission.

In this study, we revisit the A521 merging scenarios
using our WL mass and galaxy distributions in addition to the previous X-ray and radio relic observations (Figure~\ref{fig:radio_xray_mass_map}).
The comparison between galaxy and mass distributions shows that A521 has the three distinct substructures referred to as the SE, C, and NW clumps in Figure~\ref{fig:member_galaxy_with_img}. A close inspection on the galaxy distribution indicates that the C clump may be further resolved into two subclumps approximately aligned with the two X-ray peaks. On the other hand, our WL mass map reveals only a single clump in-between the two X-ray peaks. {\it If} this mass clump solely corresponds to the northern X-ray peak, the only probable mass clump associated with the southern X-ray peak is the SE clump.
Of course, the challenge with this merger scenario is the large ($\gtrsim 0.5~\rm Mpc$) dissociation distance between mass and gas, which requires a strong gas interaction
during the collision.

We perform a toy simulation of this scenario to test whether such a large gas-mass separation can occur.
Although, as mentioned above, A521 has at least three subclusters, we assume a binary collision for simplicity.
That is, we hypothesize that the NW clump does not actively participate in this merger leading to the observed dissociation and radio relic.
We setup a collision of two spherical clusters in an isolated box of $(7\rm\,Mpc)^3$.
Each cluster consists of a dark matter halo and an ICM that follow the NFW \citep{1997ApJ...490..493N} and beta profiles \citep{1976A&A....49..137C}, respectively.
Our WL mass (Table ~\ref{Tab:sub_clump_mass}) is used to setup each halo mass.
We placed these two clusters with the initial separation of $2.5\rm~Mpc$ and relative velocity of $900\rm~km~s^{-1}$. We used a small impact parameter of $200\rm~kpc$ to simulate a slight, off-axis merger.
The simulations are computed with the adaptive mesh refinement code {\tt RAMSES} \citep{Teyssier2002} and the clusters are resolved by $\mytilde100\rm~kpc$ at the outskirt and $\mytilde1\rm\,kpc$ near the center.

We convert the simulation results to observable properties including the mock X-ray image using {\tt pyXSIM}\footnote{\url{http://hea-www.cfa.harvard.edu/~jzuhone/pyxsim}} and the mock WL mass map using the projected mass distribution. The position of the merger shock is inferred by identifying a hot ($>10\rm\,keV$) region in the cluster outskirt ($>0.5\rm\,Mpc$) in the plane of the collision.
We confirm that this hot region indeed traces the merger shock by checking its motion throughout the collision history. Nevertheless, because of the low spatial resolution in the outskirts, its exact location is uncertain.

The top panel of Figure~\ref{fig:revised_scenario} provides
the schematic diagram of the merger scenario and the resulting X-ray map from the simulation. 
As the diagram describes, the head-on collision triggers the merger shock and the mass-gas dissociation.
The simulated X-ray map shows the epoch at the first apocenter where the simulated feature is analogous to that of the A521 observation.
The result shows the large separation ($\mytilde0.5\rm~Mpc$) between the two mass peaks, in a reasonable agreement with the observed separation ($\mytilde0.65\rm~Mpc$), given
the fact that we did not fine-tune the simulation setup.
Also, qualitatively, the asymmetric and elongated morphology of the X-ray emission and the close distance between the merger shock and the southern mass peak are reproduced. Nevertheless, the separation between the mass clump and the X-ray peak becomes negligibly small at this epoch. This is because the dissociation occurred during their first passage rapidly diminishes in time. Also, unlike the observation, the northern dark matter halo leads the ICM peak. 

Having seen that it is impossible to create a gas-mass dissociation as large as $\mytilde0.5$~Mpc with the first scenario, we
turn to another possibility. In this scenario, we hypothesize that the C mass clump is in fact comprised of two mass clumps. This scenario is motivated
by the features seen in our luminosity-weighted galaxy map and the {\it HST} WL mass map (right panel of Figure~\ref{fig:color_map_mass_map}). The {\it  HST} WL mass map shows that the convergence peak (where the kappa value is highest) is offset $\mytilde25\arcsec$ from the centroid defined by the large-scale mass distribution in the C clump region.  The feature indicates the possibility that there might be two overlapping mass halos in the HST WL map.
This requires an off-axis cluster merger 
and modification of the main and sub cluster masses.
We assume that the C clump is comprised of two halos with the masses of
 3.5 $\times10^{14}M_{\sun}$ and 1.5 $\times10^{14}M_{\sun}$ for the main and sub clusters, respectively\footnote{The current observation provides a constraint only for the total mass $\mytilde5\times10^{14}M_{\sun}$. Thus, the mass ratio that we employed here is somewhat arbitrary.}.
These clusters are positioned to have a similar separation as the previous scenario (2.5 Mpc) with a larger impact parameter (600 kpc), which leads to the distance $\mytilde200$ kpc at the closest passage. We assigned a lower initial velocity ($600\rm~km~s^{-1}$) because of the smaller mass.
We display this scenario in the bottom row of Figure~\ref{fig:revised_scenario}.
The interaction during the off-axis passage is much weaker than in the case of the previous head-on collision, and thus the dissociation also weakens. In the later phase, the ICM gains angular momentum and overruns the dark halo when it reaches the apocenter \citep[e.g.][]{2018ApJ...865..118S,2020ApJ...894...60L}. As a result, the ICM lags behind and features a binary distribution in X-ray, while the mass clumps start their second passage and appear as a single elongated mass map. The simulated X-ray map also reproduces the observed features such as the long and asymmetric X-ray tail of the northern cluster, the hot intermediate region due to the second in-fall, and the distant merger shock triggered by the previous collision. Consequently, our experiments with numerical simulations favor the second scenario. Nevertheless, this second scenario requires the assumption that the NW and SE mass clumps are bystanders in the merger.

Finally, one can consider a three-body collision, where all of our identified mass clumps collide and make a complex dissociative merger. Although it requires elaborate fine-tuning, this multi-body collision may explain why the NW and SE clumps are not detected in X-ray.
This numerical experiment with the three-body collision is beyond the scope of the current paper. It will certainly be interesting for future studies to find A521 analogs created by a multi-body collision in cosmological simulations.

\section{Conclusion}
\label{sec:conclusions}

The clumpy and dynamically active cluster, A521 has been studied extensively in many astrophysical contexts based on optical, X-ray, and radio observations. 
The existence of the radio relic and halo in the cluster supports its post-merger state and provides hints leading to deeper understanding of their formation mechanism during the merger.
In this study, we revisit the merging scenario of A521 based on the enhanced member catalog and improved WL analysis. 

Our enhanced cluster member catalog provides a much higher S/N galaxy distribution than the previous study and reveals that A521 is mainly composed of three distinct substructures that we refer to as C, NW, and SE.
This A521 structure is significantly simpler than the previous view that the system consists of more than seven substructures.
Our WL mass reconstruction is remarkably consistent with this simpler view of the system.

We find that the total mass of A521 is $M_{200}=13.0^{+1.0}_{-1.3}\times10^{14} M_{\sun}$, which is twice larger than the previous WL mass estimate. The large discrepancy is attributed to the difference in both treatment of the substructure and  WL analysis method.
We also estimate the masses of the individual clumps, which are used as a crucial input to our numerical simulation.

We run numerical simulations for two merging scenarios. In one scenario, the SE and C clumps experience a head-on collision. The simulation reproduces the close distance between the radio relic and the southeastern mass clump. However, with this scenario it is impossible to explain the large separation between the SE clump and the southern X-ray clump. In the other scenario, we hypothesize that the C clump seen in WL is comprised of two subclumps, as suggested
by the high-resolution {\it HST} WL.
Our off-axis collision simulation with this assumption reproduces the position and the direction of the radio relic, the X-ray morphology, and the gas-mass offset. In this case, we are witnessing the onset of the second pass-through.

We thank Giulia Macario for sharing the reduced GMRT data.
M. Yoon acknowledges support from the Max Planck Society and the Alexander von Humboldt Foundation in the framework of the Max Planck-Humboldt Research Award endowed by the Federal Ministry of Education and Research. M. Yoon also acknowledges support from the National Research Foundation of Korea (NRF) grant funded by the Korea government (MSIT) under no. 2019R1C1C1010942.
M. J. Jee acknowledges support for the current research from the National Research Foundation of Korea under the program nos. 2017R1A2B2004644 and 2017R1A4A1015178.
Hectospec observations reported here were obtained at the MMT Observatory, a joint facility of the Smithsonian Institution and the University of Arizona.
This work was supported by K-GMT Science Program of Korea Astronomy and Space Science Institute.

\bibliographystyle{apj.bst}
\bibliography{main} 
\end{document}